\pdfoutput=1
\documentclass{article}
\PassOptionsToPackage{dvipsnames}{xcolor}   %
\usepackage{swim_preprint,times}
\usepackage[T1,OT1]{fontenc}

\usepackage{amsmath,amssymb}
\usepackage{graphicx}
\usepackage[dvipsnames]{xcolor}
\usepackage{colortbl}
\usepackage{booktabs}
\usepackage{multirow}
\usepackage{pifont}
\usepackage{xspace}
\usepackage{hyperref}
\usepackage{xurl}
\usepackage{comment}
\usepackage{xspace}
\usepackage[capitalize]{cleveref}
\usepackage{wrapfig,lipsum,booktabs}

\crefname{section}{Section}{Sections}      \Crefname{section}{Section}{Sections}
\crefname{subsection}{Section}{Sections}   \Crefname{subsection}{Section}{Sections}
\crefname{table}{Table}{Tables}            \Crefname{table}{Table}{Tables}
\crefname{figure}{Figure}{Figures}         \Crefname{figure}{Figure}{Figures}
\crefname{equation}{Equation}{Equations}   \Crefname{equation}{Equation}{Equations}

\graphicspath{{Figures/}{./}}

\providecommand{\degree}{\ensuremath{^\circ}}

\makeatletter
\DeclareRobustCommand\onedot{\futurelet\@let@token\@onedot}
\def\@onedot{\ifx\@let@token.\else.\null\fi\xspace}
\def\eg{\emph{e.g}\onedot} 
\def\ie{\emph{i.e}\onedot} 
\def\etc{\emph{etc}\onedot}
\makeatother

\newcommand{\name}{SWIM\xspace}

\definecolor{best}{HTML}{D5E8D4}
\definecolor{second}{HTML}{FFF2CC}

\newcommand{\projecturl}{\url{https://binglunwang.github.io/SWIM/}}

\title{SWIM: Compact Environment Representation for Reinforcement Learning Human Swimming}

\swimfinalcopy
\author{
\begin{tabular}{@{}c@{}}
Binglun Wang$^{1}$ \quad Do\u{g}a Y\i{}lmaz$^{1}$ \quad Niloy Mitra$^{1}$ \quad Edmond S. L. Ho$^{2}$ \quad He Wang$^{1*}$ \\
{\normalfont\small $^{1}$University College London \quad $^{2}$University of Glasgow}
\end{tabular}
}

\hypersetup{pdftitle={SWIM: Compact Environment Representation for Reinforcement Learning Human Swimming},pdfauthor={Binglun Wang, Do\u{g}a Y\i{}lmaz, Niloy Mitra, Edmond S. L. Ho, He Wang}}
\begin{document}
\maketitle
\lhead{}\renewcommand{\headrulewidth}{0pt}   %
{\let\thefootnote\relax\footnotetext{$^{*}$Corresponding author.}}

\begin{abstract}
Deep reinforcement learning (RL) has driven rapid progress in physically-based motion generation, yet synthesizing robust motion policies in dense fluid environments (\eg swimming) remains unsolved. Unlike land motion, where environmental impact is sparse and can be coarsely modeled (\eg gravity, normal reaction), swimming requires continuous, full-body coordination under pressure and flow forces across the entire body surface; fully-coupled rigid–fluid simulation is far too slow for the millions of interactions RL requires.
We propose SWIM, an RL framework for physically-based human swimming learned from a single reference motion. Its core is a \textit{compact, low-dimensional environment representation:} a dual-branch graph-convolutional VQ-VAE that tokenizes per-link body–water forces and torques, informative enough for control yet robust to the rapidly changing force exchanges that would otherwise destabilize training. We pair this with a GPU-based Lagrangian solver with rigid–fluid coupling that yields 10--15$\times$ faster training.
Across hundreds of training runs and several hundred zero-shot conditions, SWIM generalizes to unseen goals and trajectories, fluids with varying density, external flows and perturbations, altered body morphologies, and all four competitive swimming styles. Against imitation-learning baselines (MimicKit-DeepMimic, MimicKit-AMP, ADD) and alternative force models (a PINN world model, MuJoCo's inertial fluid model, and an underwater-robotics simulator), SWIM achieves  better stability, goal satisfaction, and physical realism: \textit{a 63.9\% success rate on held-out generalization  versus 36.8\% for the strongest baseline}.

\end{abstract}

\begin{figure}[h!]
\centering
\includegraphics[width=\linewidth]{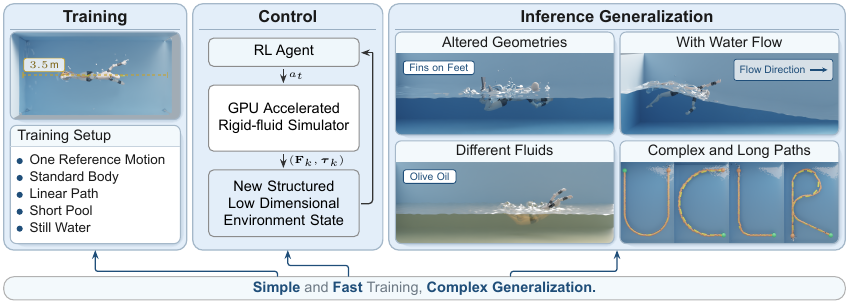}   %
\caption{Overview of SWIM. Trained on a single simple goal-reaching task in still water (left), SWIM learns a policy (middle) that can generalize to significantly different tasks, fluid environments, perturbations, body geometries, across multiple swimming styles (right).}
\label{fig:teaser}
\end{figure}

\section{Introduction}
\label{sec:intro}

Synthesizing physically-based human motion is a key challenge spanning machine learning, control theory, robotics, and now embodied AI. The primary goal is to learn robust and generalizable control policies that adhere to physical laws, while reacting naturally to environmental variations. Recently, reinforcement learning~(RL) has made impressive progress in increasingly complex environments. Existing methods can now control virtual humans/agents to perform dynamic skills, traverse uneven terrains, and interact with objects \citep{zhang_learning_2023, tessler_maskedmimic_2024, wang_skillmimic_2025, wang20261000}. Much of this success comes from advances in fast simulation, state representations, and optimized training schemes. However, these successes are predominantly confined to land scenarios where distributed environmental impact can be ignored or coarsely modeled (e.g., gravity).

We push the complexity of environments by investigating physically-based control for swimming. Swimming, by its nature, introduces a challenging paradigm of dense character-environment interaction. The (control) agent must dynamically apply forces to the environment to generate reactive propulsion and successfully swim (forward). However, unlike simple ground contact (\ie contact force) in land motion, the aquatic medium continuously applies pressure and fluid forces across the entire body surface. These environmental forces are dynamic, frequently shifting in magnitude and direction even in still water conditioned on the style/type of swimming, let alone in dynamic flows and waves. Mastering this full-body coordination requires years of training for humans (\eg to stay afloat, swim,  and/or move towards a target). Learning such a control policy requires a notable leap in handling task complexity. Previously, optimization was employed by a few for swimming control \citep{yang_layered_2004, si_realistic_2014} but merely to produce prescribed motions (\eg motion capture data). Learning control generalizable to unseen tasks/environments remains an open challenge.

The primary challenge in extending RL to swimming requires efficiently and effectively modeling dynamic environmental forces. While standard fully-coupled rigid-fluid simulations provide accurate body-water forces necessary for control learning, they are prohibitively slow for RL that requires millions of environment interactions during training \citep{gissler_interlinked_2019, takahashi2022elastomonolith, takahashi_monolith_2020, fang_iq-mpm_2020}. On the other hand, while fast surrogate models exist \citep{vaxenburg_whole-body_2025, sanchez2020learning, song_oceansim_2025, dou_cfc_2025}, they often fail to capture complex, full-body body-fluid interactions necessary for continuous swimming motion. An immediately following challenge is the high-dimensionality of the environment, \ie water, applying rapidly and drastically changing forces continuously on the body surface. Since the water-body environment must be discretized and captured in RL states for learning, suitable representations are needed to balance physical fidelity (\eg water-body force accuracy) and learning efficiency. 
Finally, similar to land motions, learning human-like motions requires data guidance as RL can `cheat' for rewards by learning unnatural motions. However, unlike land motions, high-quality full-body motion capture data for swimming remains scarce, making data-hungry imitation methods infeasible.

To address the above challenges, we propose \name (Compact Environment Representation for RL Human Swimming). Our core contribution is a novel approach to model fluid-character dynamics that strikes a balance among physical fidelity,   computational efficiency, and generalization. First, we introduce a low-dimensional state representation that effectively distills complex character-fluid interactions. This representation is sufficiently informative regarding local contact forces and the near-body fluid environment; at the same time, it is robust to the quickly changing water-body force exchanges. Second, to make training tractable, we redesign the learning loop with an optimized, GPU-based Lagrangian fluid simulator integrated with an articulated-body solver. This achieves upwards of 15$\times$ speedup in training, tackling the simulation bottleneck. Together, these force-modeling advancements enable SWIM to learn robust control policies from a single reference motion (containing only local joint kinematics) and generalize them broadly. See Figure~\ref{fig:teaser}.

We evaluate SWIM across diverse tasks, environments, and swimming styles using more than $250$ RL training runs and $80$k zero-shot generalization settings spanning over diverse environmental dimensions (\eg fluid density/viscosity, dynamic water speed/direction, external perturbations, and morphological changes). Compared to baselines adapting human imitation learning and surrogate models, SWIM demonstrates superior stability, robustness, and physical realism. 
Our key contributions are: 
(i) To the best of our knowledge, SWIM is the first method to learn stable, controllable, and natural full-body swimming policies with deep RL, from only a \emph{single} reference motion.
(ii) We introduce a structured state that tokenizes per-link hydrodynamic forces and torques via a dual-branch graph-convolutional VQ-VAE, informative enough for control yet robust to the rapidly changing force exchanges that would otherwise destabilize training. 
(iii) Across hundreds of training runs and several hundred zero-shot conditions, SWIM generalizes to unseen goals, trajectories, fluids, flows, perturbations, morphologies, and all four competitive strokes, achieving a 63.9\% success rate on held-out generalization tasks versus 36.8\% for the strongest baseline.
In addition, we also contribute a non-trivial GPU implementation of a fast and stable rigid-fluid simulator.

\section{Related Work}
\label{sec:rel}
The overall positioning of our work is shown in \Cref{tab:method_comparison}. RL-based methods have demonstrated strong performance in locomotion and agile behaviors~\citep{peng_deepmimic_2018, liu_learning_2017, yang_learning_2022}, and have been extended to more complex tasks such as sports \citep{zhang_learning_2023, wang_skillmimic_2025}, fine manipulation \citep{yang_learning_2022,music}, and large-scale motion modeling \citep{tessler_maskedmimic_2024, yao_controlvae_2022, yao_moconvq_2023}. However, they typically assume simple environments with sparse interactions (\eg motion under gravity). Swimming represents a significantly more challenging setting that involves continuous full-body interaction with a dynamic fluid environment. Early methods rely on hand-designed controllers and optimization~\citep{tan_articulated_2011, si_realistic_2014, lentine_creature_2011, kwatra_fluid_2010, yang_layered_2004}, which require extensive tuning and lack generalizability. Recent work has  started to investigate body–water interactions~\citep{dou_cfc_2025} but only on adapting motions in simple body-water interaction. Our research is the first RL-based method for physically-based character swimming motions, to the best of our knowledge.
\begin{wraptable}{r}{8.5cm}
    \centering
    \vspace{-0.5cm}
    \caption{Comparison with literature.}
    \label{tab:relate_work}
    \footnotesize
    \setlength{\tabcolsep}{3pt}
    \renewcommand{\arraystretch}{0.9}
    \begin{tabular}{lccc}
        \toprule
        Representative work & Humanoid & RL-based & Swimming \\
        \midrule
        \citet{yang_layered_2004}                              & \ding{51} & \ding{55} & \ding{51} \\
        \citet{kwatra_fluid_2010}                            & \ding{51} & \ding{55} & \ding{51} \\
       \citet{tan_articulated_2011}                            & \ding{55} & \ding{55} & \ding{51} \\
         \citet{si_realistic_2014}                                & \ding{51} & \ding{55} & \ding{51} \\
        \citet{peng_deepmimic_2018}                              & \ding{51} & \ding{51} & \ding{55} \\
        \citet{vaxenburg_whole-body_2025}                 & \ding{55} & \ding{51} & \ding{55} \\
        \citet{chu_marinegym_2025}                                   & \ding{55} & \ding{51} & \ding{51} \\
        \citet{Song2025swim} & \ding{55} & \ding{51} & \ding{51} \\
        \midrule
        \textbf{SWIM(Ours)}                                                     & \ding{51} & \ding{51} & \ding{51} \\
        \bottomrule
    \end{tabular}
    \vspace{-0.3cm}
\end{wraptable}
In fluid simulation, Eulerian methods provide high accuracy but are often prohibitively slow for RL learning \citep{takahashi_monolith_2020,takahashi2022elastomonolith, wang2024physics}. In contrast, Lagrangian approaches are fast but comparatively less accurate in computing force exchanges between rigid bodies and fluid \citep{koschier_smoothed_2019, wang2024physics}. Recently, neural network approaches \citep{jain_neural_2024, raissi_physics-informed_2019}  handle rigid-fluid coupling but their accuracy and generalization remain  challenging \citep{thuerey2021physics,mcgreivy2024weak}. In our research, Lagrangian methods provide the flexibility in balancing accuracy and speed for learning. We adopt and provide a non-trivial GPU implementation of DFSPH~\citep{bender_divergence-free_2015} with stable rigid–fluid coupling~\citep{akinci_versatile_2012}. 

Control in fluid environments has been studied in underwater robots \citep{du_underwater_2021, chen_reinforcement_2022}, marine vehicles \citep{chu_marinegym_2025, song_oceansim_2025}, and biological locomotion \citep{satoi2016unified}, as well as differentiable fluid–rigid systems~\citep{li_difffr_2023}. However, these systems typically involve low degree-of-freedom bodies or simplified interactions. In contrast, we address full-body control under complex and dynamic fluid conditions with high degree-of-freedom bodies.

\section{Methodology}
\label{sec:method}

\subsection{Problem Formulation}
\label{sec:problem}
We formulate swimming control as a Markov Decision Process $\mathcal{M} = (\mathcal{S}, \mathcal{A}, \mathcal{T}, \mathcal{R}, \pi)$. Given an initial state $s_0\in\mathcal{S}$, which describes the character, the environment, and the task, the character chooses an action $a_t \in \mathcal{A}$ at time $t$ based on a policy $\pi(a_t \mid s_t)$. During the process, the state is updated by a transition function $\mathcal{T}(s_{t+1} \mid s_t, a_t)$. At time $t$, we can evaluate a reward $r_t$ via $\mathcal{R}(a_t, s_t, s_{t+1})$. The objective is to learn the policy $\pi$ that maximizes $J(\pi) = \mathbb{E}_{\pi}\!\left[\sum_{t} \gamma^{t}\, r_t \right].$
where $\gamma\in (0,1]$ is the discount factor that determines the relative importance of immediate versus future rewards. Next, we detail $\mathcal{S}, \mathcal{A}, \mathcal{T}, \mathcal{R}, \pi$ and learning algorithm. More details are in the supplementary material.

\subsection{State Space for Body-Water Interaction}
\label{sec:state}
The state $s_t = [s_t^{char}, s_t^{goal}, s_t^{env}]$ consists of body kinematics $s_t^{char}$, the goal $s_t^{goal}$, and the surrounding environment $s_t^{env}$. Our character is an articulated humanoid with 36 controllable degrees of freedom (DoFs) and a 6-DoF root (\cref{fig:tokenizer} left). Similar to prior research~\citep{ zhang2025ADD}, $s_t^{\text{char}} \in \mathbb{R}^{174}$ includes joint position, angle, velocity, angular velocity. Also, we include two variables to parameterize stroke cycles: a phase and a frequency (as explained in \cref{sec:action}), to disambiguate recurring poses in different cycles and motion frequency. More details are in the supplementary material.

\begin{wrapfigure}{r}{7.5cm}
\centering
\vspace{-0.4cm}
\includegraphics[width=\linewidth]{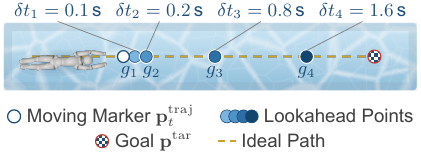}
\vspace{-0.7cm}
\caption{The goal state $s_t^{\mathrm{goal}}$ contains four lookahead waypoints $g_i(t)$ at time offsets $\delta t_i$.}
\label{fig:goal_state}
\vspace{-0.4cm}
\end{wrapfigure}

We propose a multi-horizon goal representation $s_t^{\mathrm{goal}} \in \mathbb{R}^{12}$, with an example shown in \cref{fig:goal_state}:
\begin{equation}
s_t^{\mathrm{goal}} = \left[ g_1(t)^\top, g_2(t)^\top, g_3(t)^\top, g_4(t)^\top \right]^\top \in \mathbb{R}^{12},
\end{equation}
where $g_i(t) = [x_i^{\mathrm{loc}}, y_i^{\mathrm{loc}}, \psi_i^{\mathrm{loc}}]^\top$ represents the $i$th waypoint parameterized by its location [$x_i^{\mathrm{loc}}, y_i^{\mathrm{loc}}$] in the x-y plane (z axis being the up direction) and local heading deviation $\psi_i^{\mathrm{loc}}$ toward the next waypoints at $\delta t_i \in \{0.1,\; 0.2,\; 0.8,\; 1.6\}\,\mathrm{s}$. More details are in the supplementary material. The multi-horizon goal design is for balancing motion quality and controllability. Only aiming for the final destination provides limited guidance in the middle, leading to poor controllability, while overly dense control (\eg control in every frame) is restrictive for the character to adapt to water dynamics. This is because swimming maneuvers such as steering can only be effectively executed in certain phases of a stroke cycle, \eg steering in butterfly requires asymmetrical arm pulling which can only be effective when the upper body is in water. Forcing these maneuvers by dense control at best generates unnatural motions and at worst fails to swim at all.

\subsection{Environment State}
\label{sec:envstate}
The biggest challenge of SWIM is the environment state representation which needs to balance between learning efficiency and stability, finding a sweet spot of how accurate the water-body force needs to be. Highly accurate forces require high-resolution solvers (\eg Finite Element methods with ultra-fine meshes) that can accurately compute water-body coupling, which is prohibitively slow for RL. Conversely, surrogate models such as neural solvers are sufficiently faster but often lead to inaccurate force estimation, which affects learning, as shown later. Furthermore, regardless of the solver, body-water forces change rapidly in swimming, leading to unstable training. Therefore, we propose a representation that is sufficiently accurate, fast to compute, and changes reasonably smoothly.

\begin{wrapfigure}{r}{7cm}
\centering
\vspace{-0.88cm}
\includegraphics[width=\linewidth]{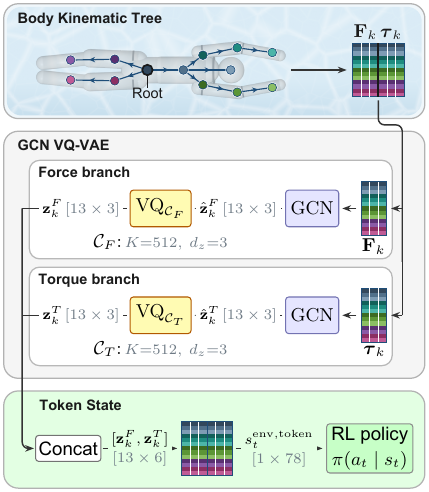}
\caption{\textbf{Dual Directed-Branch Graph Convolutional Network VQ-VAE tokenizer.} Hydrodynamic forces $\mathbf{F}_k$ and torques $\boldsymbol{\tau}_k$ propagate from parent nodes to child nodes along the skeleton. Separate codebooks quantize $\mathcal{C}_F$ and $\mathcal{C}_T$, providing a final $s_t^{\mathrm{env,token}}= [\mathbf{z}_{k}^{F},\mathbf{z}_{k}^{T}]$ for RL learning.}
\label{fig:tokenizer}
\vspace{-1.2cm}
\end{wrapfigure}

Observing that in swimming the body-water force is only qualitatively sensed by our body, not exactly computed in our brain, we abstract the force in space and time in a compact and structured. We first divide the body into 13 links (\cref{fig:tokenizer} left) and the environment state $s_t^{\mathrm{env}}=[s_t^{\mathrm{env,total}}, s_t^{\mathrm{env,token}}]$ describes the whole-body wrench $s_t^{\mathrm{env,total}}$ and individual body-water forces $s_t^{\mathrm{env,token}}$. The whole-body wrench is the net force and net torque on the whole body:
\begin{equation}
s_t^{\mathrm{env,total}} = \big[\mathbf{F}_{\mathrm{total}}^\top,\; \boldsymbol{\tau}_{\mathrm{total}}^\top\big]^\top,
\quad
\mathbf{F}_{\mathrm{total}} = \sum_{k=1}^{13} \mathbf{F}_k,
\quad
\boldsymbol{\tau}_{\mathrm{total}} = \sum_{k=1}^{13} \Big(\boldsymbol{\tau}_k + (\mathbf{c}_k - \mathbf{c}_{\mathrm{root}}) \times \mathbf{F}_k\Big),
\label{eq:env_total}
\end{equation}
where $\mathbf{c}_k, \mathbf{c}_{\mathrm{root}} \in \mathbb{R}^3$ denote the center of mass of body $k$ and the root, respectively, with $\mathbf{F}_k \in \mathbb{R}^3$ and $\boldsymbol{\tau}_k \in \mathbb{R}^3$ being the force and torque acting on body $k$ (\cref{sec:sim}). Next, the body-water forces are also represented for each body link as a compact representation of fluid interaction:
\begin{equation}
s_t^{\mathrm{env,token}} = \big[\mathbf{z}^{F}_{1,t}{}^\top,\; \mathbf{z}^{T}_{1,t}{}^\top,\; \ldots,\; \mathbf{z}^{F}_{13,t}{}^\top,\; \mathbf{z}^{T}_{13,t}{}^\top\big]^\top \in \mathbb{R}^{78},
\label{eq:env_token}
\end{equation}
where $\mathbf{z}^{F}_{k,t}, \mathbf{z}^{T}_{k,t} \in \mathbb{R}^{d_z}$ are features representing the force and torque applied to body $k$ by water. Note that $\mathbf{z}^{F}_{k,t}, \mathbf{z}^{T}_{k,t}$ are not raw forces but qualitative representations of body-water forces. For the considerations explained above, these representations need to mimic how humans feel (\ie qualitative sensing) water dynamics around different body parts when swimming. In addition, they need to faithfully represent the raw forces but they themselves should not change too abruptly in time. To this end, in practice, we propose to tokenize individual body-water forces, via a dual-branch graph convolutional vector quantizer based on VQ-VAE \citep{van2017neural}, shown in \cref{fig:tokenizer}. Our limb-level tokenizer is built around two considerations. First, force and torque cause two types of motions in different spaces: linear and angular motions. We therefore employ two sub-networks with independent codebooks ($\mathcal{C}_F$ and $\mathcal{C}_T$, each containing $K=512$ vectors of dimension $d_z=3$), to capture quantized force and torque separately in a compact and stable form. 

Second, both force and torque propagate on skeleton following a tree structure which is a directed graph from parent joints to child joints. This is essential as the dynamics of the parent link largely determines that of the child link. As shown in the supplementary material, this one-way propagation is crucial in learning and zero-shot transfer. Specifically, the encoder (\cref{fig:tokenizer} middle) maps raw forces and torques of individual bodies to latents, which are then quantized using their respective codebooks:
\begin{equation}
\mathbf{z}_{k}^{F} = \operatorname{VQ}_{\mathcal{C}_F}(\hat{\mathbf{z}}_{k}^{F}),
\quad
\mathbf{z}_{k}^{T} = \operatorname{VQ}_{\mathcal{C}_T}(\hat{\mathbf{z}}_{k}^{T}).
\label{eq:vq_lookup}
\end{equation}
The words $\mathbf{z}_{k}^{F}$ and $\mathbf{z}_{k}^{T}$ are latent representations of water-body forces. We pre-train the encoder using self-supervised wrench reconstruction and freeze it during RL learning, so the control is learned in the latent space, while preserving the key physical information of the original wrench.

\subsection{Policy and Action Space}
\label{sec:action}
The policy $\pi$ is a neural network that maps a state $s_t$ to an action $a_t$: $a_t = [\omega_t, \Delta\mathbf{q}_t] \in \mathbb{R}^{1+36}$, where $\omega_t$ is stroke frequency and $\Delta\mathbf{q}_t \in \mathbb{R}^{36}$ is a correction ~\citep{bergamin2019drecon, yuan2020residual}, both of which are used to compute the target of the control:
\begin{equation}
\begin{aligned}
&\phi_{t} = \big(\phi_{t-1} + \omega_t \Delta t\big) \in [0, 1),\ \quad \\
&\mathbf{q}_t^{\mathrm{target}} = \mathbf{q}^{\mathrm{ref}}(\phi_t) + \Delta\mathbf{q}_t,
\end{aligned}
\label{eq:policy}
\end{equation}
where $\Delta t$ is the time step, $\mathbf{q}_t^{\mathrm{target}}$ is the target pose for proportional-derivative (PD) control \citep{peng2025mimickit}, $\phi_t$ is the phase, and $\mathbf{q}^{\mathrm{ref}}(\phi_t)$ the pose sampled at the $\phi_t$ from a reference motion \Cref{sec:exp_setup}. To prevent abrupt changes, both action outputs are smoothed using first-order exponential moving average filters, with the smoothing parameter based on research on the frequency of muscle control \citep{crenna_filtering_2021}. The filtering is constrained within $\omega_t \in [0.5, 1.5]$. More details in the supplementary material.

\subsection{Body-fluid Simulation as Transition Function}
\label{sec:sim}
The transition function $\mathcal{T}(s_{t+1} \mid s_t, a_t)$ is realized as a physical simulator. We choose a Lagrangian method. Despite their general inferiority in accuracy to Eulerian methods, they strike a good balance between accuracy and speed. However, empirically we found many existing implementations are still slow. Therefore, we re-implemented the CPU-based SPlisHSPlasH simulator ~\citep{SPlisHSPlasH_Library} with DFSPH~\citep{bender_divergence-free_2015} and rigid-fluid coupling~\citep{akinci_versatile_2012} fully on GPU and gained 15$\times$ acceleration on a single NVIDIA GeForce RTX 5090 GPU. More details are in the supplementary material.

\subsection{Swimming-Informed Reward Design}
\label{sec:reward}

Our reward function $r_t = \mathcal{R}(a_t, s_t, s_{t+1})$ drives the character towards the goal by a task reward $r_t^{task}$, maintains body stability by a body stability reward $r_t^{stab}$, and preserves motion naturalness by a motion smoothness reward $r_t^{accel}$. Together, these terms provide complementary training signals. Similar to previous work \citep{mysore2021train}, we use a multiplicative-additive form: 
\begin{equation}
    r_t = r_t^{\text{task}} \cdot r_t^{\text{stab}} + w_{\text{accel}} \cdot r_t^{\text{accel}}.
    \label{eq:total_reward}
\end{equation}

\begin{figure}[b]
\centering
\vspace{-0.5cm}
\includegraphics[width=0.95\textwidth]{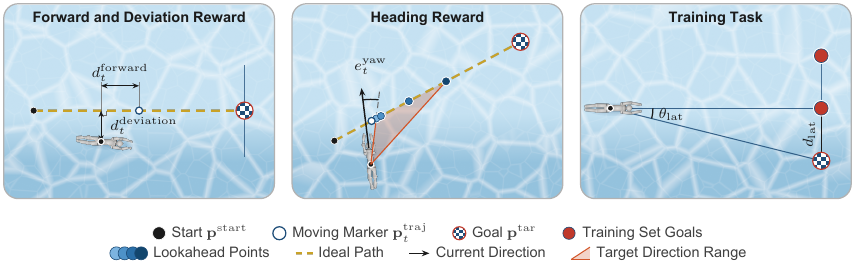}
\vspace{-0.3cm}
\caption{Left: forward and deviation reward. Middle: head reward. Right: training task. Red dots are goals. Blue dots are intermediate goals. Black dots are the initial position. The straight line between the initial position and a goal is the ideal trajectory.}
\label{fig:reward_explain}
\vspace{-0.3cm}
\end{figure}
We decompose the task into three complementary objectives. Given a target where we assume a straight path, or an arbitrary control trajectory, we consider the progress accomplished along the path, the lateral alignment of the body with the path, and the heading direction, as defined:
\begin{equation}
    r_t^{\text{task}} = r_t^{\text{forward}} \cdot r_t^{\text{deviation}} \cdot r_t^{\text{heading}} \nonumber 
    =\exp
    \left(
    -\alpha_p d_t^{\text{forward}}
    \right)
    \exp
    \left(
    -\alpha_a d_t^{\text{deviation}}
    \right)
    \exp
    \left(
    -\alpha_h e_t^{\text{yaw}}
    \right)
    \label{eq:task_reward}
\end{equation}

The $d_t^{\text{forward}}$, $d_t^{\text{deviation}}$, and $e_t^{\text{yaw}}$ denote the forward distance error, lateral deviation distance error, and heading-yaw error, respectively, as shown in \cref{fig:reward_explain}, detailed in the supplementary material.

Swimming styles such as \texttt{freestyle} require large body roll \citep{psycharakis2010body}, which during learning can easily cause unnatural motions, even flipping, of the body. To mitigate this, we design a stability reward:
$ r_t^{\text{stab}} = \exp\left(-\alpha_r \cdot d_t^{\text{roll}}\right)$. Where $d_t^{\text{roll}}$ measures the error between the current body roll angle and the ideal roll angle (\citep{psycharakis2010body}). Following \citep{peng_deepmimic_2018}, we penalize the mean absolute joint acceleration to encourage temporally smooth motion and suppress jitter. Detailed formulations are provided in the supplementary material.

\noindent
\textbf{Training design for single-instance imitation.} We design only one goal-reaching task for training explained later, to avoid long training time due to the required physical simulation. Also, we do not have reward terms requiring the generated motions to strictly follow the reference motion. This is important as we only use one reference motion for learning, which does not provide sufficient samples to describe the feasible swimming motion space for the policy. Therefore, we only restrict the updates in \cref{eq:policy} and encourage similar dynamic profiles in $r_t^{\text{accel}}$. 

\noindent
\textbf{Training.} We employ Proximal Policy Optimization (PPO) \citep{schulman_proximal_2017} for learning with several swimming-specific adaptions. More details are provided in the supplementary material.

\section{Experimental Results}
\label{sec:results}
Due to space limit of the main paper, we provide the overall evaluation settings and the main results here, and leave the full details in the supplementary material. Video demos are available on the project page (\projecturl), where code, data and supplementary material will also be released. 

\subsection{Experimental Setup}
\label{sec:exp_setup}
Since SWIM is the first RL-based method for human swimming and has no existing benchmarking protocols to follow, we first define tasks, metrics, and training and generalization settings. 

\noindent \textbf{Task.} We define two tasks: \emph{goal-reaching} and \emph{trajectory-following}. Goal-reaching is, given a starting configuration and a goal position, to test if the character can reach the goal at the specified time. Trajectory-following is to test if the character can follow a pre-defined trajectory. 

\noindent \textbf{Metrics.} We use eight metrics into two groups for task completion and motion quality, respectively. The first group includes success rate (\textbf{SR}), survival rate (\textbf{SRV}), progress (\textbf{Prog}), positional error (\textbf{PS}), and lateral deviation (\textbf{LD}): \textbf{SR} measures if the task is successful in given time; \textbf{SRV} measures if character survives testing (\eg not excessive rolling beyond $90 \degree$ or not falling out of the pool); \textbf{Prog} measures the overall progress accomplished; \textbf{PS} measures the deviation of the character from a desired trajectory; \textbf{LD} measures lateral deviation from the desired trajectory. The desired trajectory for both \textbf{PS} and \textbf{LD} is the straight line from the initial root position to the goal in goal-reaching, and the pre-defined trajectory in trajectory-following. More details are provided in the supplementary material. 
The second group includes \textbf{Roll}, \textbf{Vel} and \textbf{Accel}: \textbf{Roll} measures if body rolling exceeds the specified tolerance range over time. \textbf{Vel} and \textbf{Accel} measure the joint velocity difference between the generated and reference motion, and the joint acceleration of the generated motion itself, respectively. The definitions and justifications of the metrics are given in the supplementary material. All comparison and ablation results are averaged over three runs under the same conditions for completeness and consistency.

\noindent \textbf{Data.} The only available full-body swimming motion dataset captured from a real human is the \emph{Swimming Techniques} capture of the Porto Biomechanics Laboratory~\citep{labiomep_qualisys_capture}, a small dataset containing one instance for each style: \texttt{front-crawl}, \texttt{backstroke}, \texttt{breaststroke}, and \texttt{butterfly}, lasting only 6.44~s with only local joint-angle motion after preprocessing work. The body has 12 joints and the motion is sampled at 100Hz. During simulation, we divide the body surfaces into patches for water-body force estimation. Details are in the supplementary material.

\noindent
\textbf{Training \& Generalization.} For each swimming style, we train one policy using a simple goal-reaching task with only 3 goals spanning $\theta_{lat}=14^\circ$ in a small pool ($3.5\text{m} \times 1.8\text{m} \times 0.6\text{m}$) in still water, shown in \cref{fig:reward_explain} Right. For generalization tests, we use combinations of unseen environments including large pools ($5.0\text{m} \times 7.5 \text{m}\times 0.6$\,m and $8.0\text{m} \times 8.5\text{m} \times 0.6$\,m), goals, control trajectories, water flows, perturbations, body geometries and morphologies, and different fluids across a broad range of conditions. Note this is a strong generalization test as the training only involves short-distance swimming without control trajectories and in still water. Details are in the supplementary material.

\subsection{Strong Generalization to Unseen Conditions}
\label{sec:generalisation}

\begin{figure}[tb]
\centering
\includegraphics[width=\linewidth]{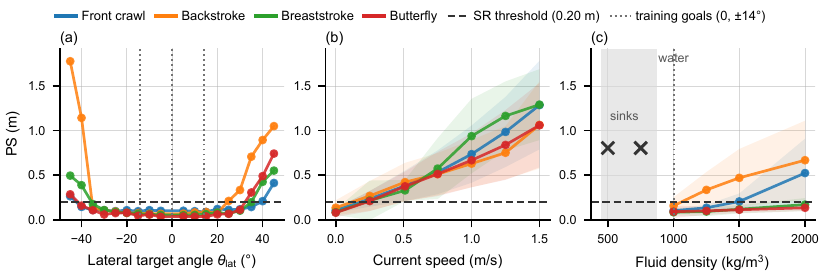}
\vspace{-0.7cm}
\caption{Position errors in different conditions. \textbf{(a)} Goal-reaching in the 5\,m pool against the lateral target angle $\theta_{lat}$; the dotted vertical lines are the three training goal angles ($0$ and $\pm14^\circ$). \textbf{(b)} Against the flow speed and averaged over four flow directions. \textbf{(c)} Against the fluid density; the dotted line is water, the training fluid. The character sinks in the shaded range (sinks) and hence no position error reported. In all figures, the horizontal dashed line is the SR threshold ($\text{PS} \le 0.20$\,m).}
\vspace{-0.5cm}
\label{fig:gen_boundary}
\end{figure}

We first report generalization results on exhaustive and challenging combinations of conditions in tasks, environments, perturbation, and body geometry and morphologies:
\begin{itemize}
    \item (\textbf{T}) 10 sampled unseen goals up to $\theta_{lat}=30^\circ$ ($2\times$ wider than training), for goal-reaching; 20 unseen control trajectories containing short quick turns and slow long turns up to $30^\circ$ represented by polylines and sinusoidal curves, for trajectory-following. A total of 30 tasks.
    \item (\textbf{E}) Water flows in 4 different directions, with 2 speeds ($\leq$ 0.51\,m/s), 3 densities (up to $1.5\times$ of water), and 5 viscosities (up to $7{,}000\times$ of water). A total of 120 conditions.
    \item (\textbf{P}) Impulse or random perturbations applied to the body, including short and strong forces, and long and light forces, varying in directions and magnitudes, providing 91 conditions.
    \item (\textbf{B}) 3 body geometries where we attach fins of 10cm, 15cm, and 20cm long to the feet, and 8 body morphologies by removing body parts. A total of 11 conditions.
    \item (\textbf{S}) 4 swimming styles: front-crawl, backstroke, breaststroke, and butterfly.
\end{itemize}
Theoretically, there are a total of 14,414,400 combinations, of which we choose 16,632 (see the supplementary material). We report the results in \Cref{tab:main_gen} across all four swimming styles across \textbf{T}$\times$\textbf{E}. Overall, SWIM achieves $\ge 98\%$ SRV and $\le0.4$ PS, showing that it can successfully exploit meaningful policies within valid physical spaces and motions, and accomplish tasks across various unseen conditions that are drastically different from that of training. This demonstrates robust task completion beyond the training distribution. Note that we have a strict \textbf{SR} criterion (detailed in \cref{fig:gen_boundary}). Despite \textbf{SR} can be as low as 27.5\%, the \textbf{Prog} shows SWIM has reasonably accomplished all tasks.
\begin{table}[tb]
\centering
\vspace{-1cm}
\caption{Results averaged across T$\times$E conditions.}
\small
\setlength{\tabcolsep}{2pt}
\resizebox{\textwidth}{!}{%
\begin{tabular}{ll|rrrrrrrr}
\toprule
Task & Stroke & SRV$\uparrow$ (\%) & SR$\uparrow$ (\%) & PS$\downarrow$ (m) & LD$\downarrow$ (m) & Prog$\to1$ & Roll$\downarrow$ ($^\circ$) & Vel$\downarrow$ (rad/s) & Accel$\downarrow$ (rad/s$^2$) \\
\midrule
Goal reaching & Front crawl & \ensuremath{100.0} & \ensuremath{58.8} & \ensuremath{0.223_{\pm0.152}} & \ensuremath{0.102_{\pm0.113}} & \ensuremath{0.856_{\pm0.088}} & \ensuremath{1.02_{\pm0.88}} & \ensuremath{1.04_{\pm0.09}} & \ensuremath{54.92_{\pm3.27}} \\
 & Backstroke & \ensuremath{99.4} & \ensuremath{32.5} & \ensuremath{0.387_{\pm0.331}} & \ensuremath{0.198_{\pm0.201}} & \ensuremath{0.805_{\pm0.180}} & \ensuremath{2.00_{\pm2.65}} & \ensuremath{1.19_{\pm0.04}} & \ensuremath{53.49_{\pm2.23}} \\
 & Breaststroke & \ensuremath{100.0} & \ensuremath{62.5} & \ensuremath{0.211_{\pm0.221}} & \ensuremath{0.098_{\pm0.166}} & \ensuremath{0.878_{\pm0.123}} & \ensuremath{0.82_{\pm0.72}} & \ensuremath{1.65_{\pm0.08}} & \ensuremath{95.87_{\pm10.04}} \\
 & Butterfly & \ensuremath{100.0} & \ensuremath{63.1} & \ensuremath{0.195_{\pm0.152}} & \ensuremath{0.089_{\pm0.117}} & \ensuremath{0.877_{\pm0.095}} & \ensuremath{0.54_{\pm0.46}} & \ensuremath{1.58_{\pm0.10}} & \ensuremath{76.82_{\pm6.00}} \\
\midrule
Trajectory following & Front crawl & \ensuremath{99.4} & \ensuremath{49.4} & \ensuremath{0.235_{\pm0.139}} & \ensuremath{0.122_{\pm0.119}} & \ensuremath{0.868_{\pm0.080}} & \ensuremath{1.09_{\pm1.85}} & \ensuremath{1.04_{\pm0.08}} & \ensuremath{54.40_{\pm3.21}} \\
 & Backstroke & \ensuremath{99.7} & \ensuremath{27.5} & \ensuremath{0.321_{\pm0.198}} & \ensuremath{0.191_{\pm0.143}} & \ensuremath{0.863_{\pm0.145}} & \ensuremath{2.25_{\pm3.23}} & \ensuremath{1.18_{\pm0.04}} & \ensuremath{53.08_{\pm2.47}} \\
 & Breaststroke & \ensuremath{99.1} & \ensuremath{54.7} & \ensuremath{0.210_{\pm0.128}} & \ensuremath{0.109_{\pm0.106}} & \ensuremath{0.911_{\pm0.100}} & \ensuremath{0.97_{\pm2.00}} & \ensuremath{1.65_{\pm0.08}} & \ensuremath{96.67_{\pm9.83}} \\
 & Butterfly & \ensuremath{100.0} & \ensuremath{59.1} & \ensuremath{0.215_{\pm0.154}} & \ensuremath{0.112_{\pm0.127}} & \ensuremath{0.890_{\pm0.083}} & \ensuremath{0.64_{\pm0.72}} & \ensuremath{1.57_{\pm0.09}} & \ensuremath{76.65_{\pm5.51}} \\
\bottomrule
\end{tabular}}%
\vspace{-0.4cm}
\label{tab:main_gen}
\end{table}

In goal-reaching, \Cref{fig:gen_boundary} shows that SWIM generalizes up to $\theta_{lat}=30^\circ$ while maintaining a position error below $0.2$ m. Beyond $\theta_{lat}=30^\circ$ in both tasks, flow speed larger than $0.51m/s$, and extreme fluid densities ($1500 kg/m^3$), SWIM starts to fail more often. However, we argue these combinations are challenging for humans too, \eg $0.51 m/s$ is a dangerous flow for swimming already \citep{national2004nfpa}, highlighting the difficulty of these unseen conditions in practice.

\begin{wrapfigure}{r}{8cm}
\centering
\vspace{-0.3cm}
\includegraphics[width=\linewidth]
{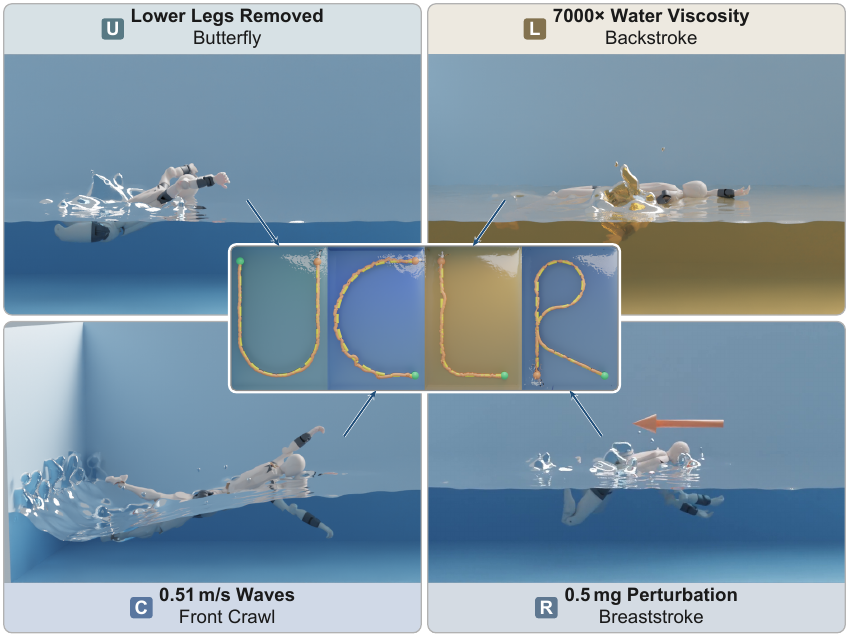}
\vspace{-0.7cm}
\caption{`UCLR' trajectory-following under challenging conditions. Orange curves are from synthesized motions.}
\label{fig:uclr_word_still}
\vspace{-0.3cm}
\end{wrapfigure}

We show one example of trajectory-following in \Cref{fig:uclr_word_still}, which shows SWIM can follow complex trajectories in unseen conditions, even only trained on a simple goal-reaching task. More quantitative and visual results in the supplementary material  further confirm that SWIM has a strong generalization and produces natural-looking motions (see videos) across a wide range of unseen conditions.

\subsection{Comparisons}
\label{sec:comparisons}
As there are no baseline methods for direct comparison, we adapt several baseline methods in imitation learning for human motions: DeepMimic~\citep{peng_deepmimic_2018, peng2025mimickit}, AMP~\citep{peng_amp_2021, peng2025mimickit} and ADD~\citep{zhang2025ADD}. We also use a simple PD controller to follow the reference motion (data replay) as a baseline. This comparison shows SWIM is genuinely the first RL method capable for synthesizing stable, controllable and natural swimming motions. 
Furthermore, we compare SWIM against three surrogate models in the transition function realization: a world model in CFC~\citep{dou_cfc_2025} based on Physics-inspired Neural Networks (PINNs), MuJoCo's inertial fluid model~\citep{todorov2012mujoco,vaxenburg_whole-body_2025}, and OceanSim~\citep{song_oceansim_2025} designed for underwater robots. In addition, we use a high-resolution SPH simulator as another baseline which estimates water-body forces more accurately than SWIM. Overall, these baselines show that SWIM strikes a balance between accuracy and speed for effective RL learning. More details are provided in the supplementary material.

Results on \textbf{T}$\times$\textbf{E} conditions averaged over all swimming styles are shown in \Cref{tab:baselines}. First, SWIM achieves the highest on key task-related metrics such as \textbf{SR} and \textbf{Prog} by large margin. Although AMP and DeepMimic occasionally achieve better results on task metrics, their motions are unnatural, with the arms moving in bird-like flapping motions (see videos). Overall, these results show that existing methods cannot be easily adapted to the SWIM task. 

Also, \Cref{tab:baselines} bottom shows that both surrogate models and highly accurate simulators achieve low SRs (below $10\%$), with varying speeds in generating samples for RL via simulation (samples/s). SWIM is neither the fastest nor the slowest, but is the best balanced one. Surrogate models are overly inaccurate for RL learning, \eg MuJoCo not modeling dynamic water-air boundaries, while high-res simulators (High-resolution SPH) is too slow for RL training in practice. 

Finally, we further compare our GPU-based SPH simulator with the original CPU-based SPlisHSPlasH \citep{SPlisHSPlasH_Library}. On the same machine, our simulator can run 18 environments in parallel and achieves $1{,}023$ samples/s, compared with merely $65$ samples/s using 24 core CPU AMD Ryzen Threadripper 9960X CPU with SPlisHSPlasH (more details in the supplementary material). This corresponds to a higher sampling throughput of more than $15\times$. In general, these results demonstrate that SWIM achieves a favorable balance between physical accuracy and training efficiency. 
 
\begin{table}[t]
\centering
\small
\setlength{\tabcolsep}{4pt}
\caption{\textbf{Comparison after training on 10M samples.} Top: Comparison with imitation learning methods; Bottom: Comparison with surrogate models. $^{\dagger}$ trained for the same amount of time as SWIM or otherwise too slow if trained on 10M samples. The top-3 is labeled as \colorbox{blue!32}{1st} / \colorbox{blue!18}{2nd} / \colorbox{blue!8}{3rd}.}
\label{tab:baselines}
\vspace{0.1cm}
\resizebox{\textwidth}{!}{%
\begin{tabular}{l|rrrrrrrr|r}
\toprule
Method & SRV$\uparrow$ (\%) & SR$\uparrow$ (\%) & PS$\downarrow$ (m) & LD$\downarrow$ (m) & Prog$\to1$ & Roll$\downarrow$ ($^\circ$) & Vel$\downarrow$ (rad/s) & Accel$\downarrow$ (rad/s$^2$) & Samples/s$\uparrow$ \\
\midrule
\textbf{SWIM} (ours) & \cellcolor{blue!32}\ensuremath{99.7_{\pm0.2}} & \cellcolor{blue!32}\ensuremath{63.9_{\pm4.7}} & \cellcolor{blue!32}\ensuremath{0.211_{\pm0.024}} & \cellcolor{blue!8}\ensuremath{0.101_{\pm0.024}} & \cellcolor{blue!32}\ensuremath{0.857_{\pm0.015}} & \cellcolor{blue!18}\ensuremath{0.94_{\pm0.54}} & \cellcolor{blue!32}\ensuremath{1.31_{\pm0.24}} & \cellcolor{blue!32}\ensuremath{64.84_{\pm14.59}} & \ensuremath{1{,}023} \\
\midrule
AMP & \cellcolor{blue!8}\ensuremath{97.5_{\pm2.1}} & \cellcolor{blue!18}\ensuremath{36.8_{\pm44.6}} & \cellcolor{blue!18}\ensuremath{0.961_{\pm0.801}} & \cellcolor{blue!32}\ensuremath{0.089_{\pm0.032}} & \cellcolor{blue!18}\ensuremath{0.413_{\pm0.499}} & \cellcolor{blue!32}\ensuremath{0.45_{\pm0.46}} & \cellcolor{blue!18}\ensuremath{2.21_{\pm0.76}} & \cellcolor{blue!18}\ensuremath{128.67_{\pm58.98}} &  \ensuremath{940}\\
DeepMimic & \cellcolor{blue!18}\ensuremath{98.7_{\pm2.3}} & \cellcolor{blue!8}\ensuremath{12.7_{\pm25.4}} & \cellcolor{blue!8}\ensuremath{1.236_{\pm0.464}} & \cellcolor{blue!18}\ensuremath{0.099_{\pm0.023}} & \cellcolor{blue!8}\ensuremath{0.219_{\pm0.276}} & \cellcolor{blue!8}\ensuremath{3.35_{\pm4.04}} & \cellcolor{blue!8}\ensuremath{2.76_{\pm0.85}} & \cellcolor{blue!8}\ensuremath{323.75_{\pm153.02}} &  
\ensuremath{903}\\
ADD & \ensuremath{98.4_{\pm2.5}} & \ensuremath{0.0_{\pm0.0}} & \ensuremath{1.597_{\pm0.138}} & \ensuremath{0.122_{\pm0.054}} & \ensuremath{0.028_{\pm0.081}} & \ensuremath{0.46_{\pm0.63}} & \ensuremath{1.65_{\pm0.21}} & \ensuremath{107.37_{\pm27.46}} &  \ensuremath{899}\\
Data Replay & \ensuremath{100.0_{\pm0.0}} & \ensuremath{0.0_{\pm0.0}} & \ensuremath{0.900_{\pm0.278}} & \ensuremath{0.264_{\pm0.131}} & \ensuremath{0.403_{\pm0.172}} & \ensuremath{3.36_{\pm4.98}} & \ensuremath{0.73_{\pm0.21}} & \ensuremath{35.98_{\pm11.21}} &  \ensuremath{1{,}160}\\
\midrule
MuJoCo& \cellcolor{blue!18}\ensuremath{58.4_{\pm40.8}} & \cellcolor{blue!18}\ensuremath{1.8_{\pm3.5}} & \cellcolor{blue!18}\ensuremath{0.808_{\pm0.137}} & \cellcolor{blue!18}\ensuremath{0.359_{\pm0.182}} & \cellcolor{blue!18}\ensuremath{0.644_{\pm0.184}} & \cellcolor{blue!18}\ensuremath{25.12_{\pm23.24}} & \cellcolor{blue!18}\ensuremath{1.59_{\pm0.16}} & \cellcolor{blue!18}\ensuremath{81.42_{\pm17.15}} & \ensuremath{2{,}194} \\
CFC$^{\dagger}$ & \ensuremath{87.7_{\pm16.9}} & \ensuremath{0.0_{\pm0.0}} & \ensuremath{1.133_{\pm0.189}} & \ensuremath{0.205_{\pm0.099}} & \ensuremath{0.267_{\pm0.117}} & \ensuremath{10.61_{\pm9.76}} & \ensuremath{0.90_{\pm0.20}} & \ensuremath{53.03_{\pm17.37}} & \ensuremath{55} \\
OceanSim & \ensuremath{68.2_{\pm43.8}} & \ensuremath{0.0_{\pm0.0}} & \ensuremath{1.237_{\pm0.266}} & \ensuremath{0.328_{\pm0.136}} & \ensuremath{0.251_{\pm0.129}} & \ensuremath{15.68_{\pm23.35}} & \ensuremath{1.20_{\pm0.23}} & \ensuremath{67.30_{\pm18.62}} & \ensuremath{2{,}657} \\
High-res SPH$^{\dagger}$ & \ensuremath{74.6_{\pm18.1}} & \ensuremath{0.0_{\pm0.0}} & \ensuremath{1.266_{\pm0.157}} & \ensuremath{0.232_{\pm0.110}} & \ensuremath{0.204_{\pm0.086}} & \ensuremath{18.02_{\pm14.83}} & \ensuremath{1.11_{\pm0.18}} & \ensuremath{61.76_{\pm20.02}} & \ensuremath{5} \\
\bottomrule
\end{tabular}}%
\label{tab:method_comparison}
\vspace{-0.2cm}
\vspace{-0.2cm}
\end{table}

\subsection{Ablation Study}
\label{sec:ablation}
We test alternative environment representations in two dimensions: \emph{spatial resolution}, the granularity at which water-body forces are represented: whole-body, per-body, or body-surface patches (52 patches in total), and \emph{feature representation}, which specifies whether the water-body forces are represented as raw forces or tokens. Additionally, we test a feature combination of whole-body raw forces + per-body token (SWIM). Finally, we test densely sampled points on the body surface over time, forming a high-dimensional, unstructured state (High-dim state). We show one ablation study in \cref{tab:ablation_overall} under \textbf{T}$\times$\textbf{E} conditions averaged over all swimming styles here and additional ablation study are provided in the supplementary material. The detailed settings are provided in the supplementary material.

Overall, most high-dimensional representations (\eg Body-surface-patch raw/token, High-dim state) and unstructured representations (\eg Whole-body token, Per-body raw) perform badly. Tokenization is useful particularly at per-body level, \eg Per-body token. Whole-body raw and Per-body token are overall winners but the former produces less natural motions than the latter. Therefore, we choose Whole-body raw + Per-body token for SWIM, which combines an overall wrench plus quantized per-body water-body forces as a balance between tasks and motion quality.

\begin{table}[t]
\centering
\small
\setlength{\tabcolsep}{4pt}
\caption{\textbf{Environment state ablation.} The top-3 per column is labeled as \colorbox{blue!32}{1st} / \colorbox{blue!18}{2nd} / \colorbox{blue!8}{3rd}.}
\label{tab:ablation_overall}
\vspace{0.1cm}
\resizebox{\textwidth}{!}{%
\begin{tabular}{l|rrrrrrrr}
\toprule
Environment state & SRV$\uparrow$ (\%) & SR$\uparrow$ (\%) & PS$\downarrow$ (m) & LD$\downarrow$ (m) & Prog$\to1$ & Roll$\downarrow$ ($^\circ$) & Vel$\downarrow$ (rad/s) & Accel$\downarrow$ (rad/s$^2$) \\
\midrule
\textbf{SWIM} & \cellcolor{blue!32}\ensuremath{99.7_{\pm0.2}} & \ensuremath{63.9_{\pm4.7}} & \ensuremath{0.211_{\pm0.024}} & \cellcolor{blue!18}\ensuremath{0.101_{\pm0.024}} & \ensuremath{0.857_{\pm0.015}} & \cellcolor{blue!18}\ensuremath{0.94_{\pm0.54}} & \cellcolor{blue!18}\ensuremath{1.31_{\pm0.24}} & \cellcolor{blue!32}\ensuremath{64.84_{\pm14.59}} \\
Whole-body token & \cellcolor{blue!8}\ensuremath{98.2_{\pm3.4}} & \ensuremath{63.8_{\pm6.1}} & \ensuremath{0.210_{\pm0.055}} & \ensuremath{0.104_{\pm0.050}} & \cellcolor{blue!8}\ensuremath{0.874_{\pm0.015}} & \ensuremath{1.64_{\pm2.01}} & \ensuremath{1.34_{\pm0.20}} & \cellcolor{blue!18}\ensuremath{66.35_{\pm13.17}} \\
Per-body token & \ensuremath{96.4_{\pm6.8}} & \cellcolor{blue!18}\ensuremath{68.5_{\pm4.6}} & \cellcolor{blue!18}\ensuremath{0.193_{\pm0.022}} & \cellcolor{blue!18}\ensuremath{0.101_{\pm0.025}} & \ensuremath{0.871_{\pm0.015}} & \ensuremath{2.46_{\pm3.28}} & \ensuremath{1.34_{\pm0.25}} & \cellcolor{blue!8}\ensuremath{67.23_{\pm15.13}} \\
Body-surface-patch token & \cellcolor{blue!8}\ensuremath{98.2_{\pm3.4}} & \ensuremath{56.3_{\pm7.9}} & \ensuremath{0.232_{\pm0.030}} & \ensuremath{0.121_{\pm0.019}} & \ensuremath{0.850_{\pm0.024}} & \cellcolor{blue!8}\ensuremath{1.52_{\pm1.62}} & \ensuremath{1.36_{\pm0.26}} & \ensuremath{67.71_{\pm16.66}} \\
Whole-body raw & \cellcolor{blue!18}\ensuremath{99.6_{\pm0.7}} & \cellcolor{blue!32}\ensuremath{69.8_{\pm3.6}} & \cellcolor{blue!32}\ensuremath{0.177_{\pm0.018}} & \cellcolor{blue!32}\ensuremath{0.087_{\pm0.015}} & \cellcolor{blue!32}\ensuremath{0.896_{\pm0.013}} & \cellcolor{blue!32}\ensuremath{0.74_{\pm0.76}} & \ensuremath{1.36_{\pm0.20}} & \ensuremath{67.82_{\pm12.60}} \\
Per-body raw & \ensuremath{95.6_{\pm8.5}} & \cellcolor{blue!8}\ensuremath{66.8_{\pm6.6}} & \cellcolor{blue!8}\ensuremath{0.207_{\pm0.048}} & \ensuremath{0.112_{\pm0.034}} & \cellcolor{blue!18}\ensuremath{0.876_{\pm0.015}} & \ensuremath{2.93_{\pm4.69}} & \cellcolor{blue!18}\ensuremath{1.31_{\pm0.20}} & \ensuremath{71.95_{\pm16.12}} \\
Body-surface-patch raw & \ensuremath{97.7_{\pm4.1}} & \ensuremath{59.2_{\pm9.7}} & \ensuremath{0.228_{\pm0.065}} & \ensuremath{0.114_{\pm0.054}} & \ensuremath{0.847_{\pm0.010}} & \ensuremath{1.79_{\pm2.30}} & \cellcolor{blue!32}\ensuremath{1.28_{\pm0.25}} & \ensuremath{73.30_{\pm17.01}} \\
High-dim state & \ensuremath{86.6_{\pm26.9}} & \ensuremath{0.0_{\pm0.0}} & \ensuremath{1.067_{\pm0.173}} & \ensuremath{0.206_{\pm0.044}} & \ensuremath{0.308_{\pm0.110}} & \ensuremath{4.77_{\pm7.11}} & \ensuremath{0.87_{\pm0.22}} & \ensuremath{43.63_{\pm7.60}} \\
\bottomrule
\end{tabular}}%
\vspace{-0.2cm}
\vspace{-0.5cm}
\end{table}

\section{Conclusion, Limitation, and Future Work}
\label{sec:discussion}

We proposed SWIM, the first RL-based method for physically-based character swimming. By combining a structured low-dimensional environment representation and an efficient simulator, SWIM learns natural, controllable, and generalizable swimming control from a single motion. Extensive experiments demonstrate robust performance across tasks, environments, body geometries and morphologies, and swimming styles under a wide range of unseen conditions. 

One limitation is that SWIM relies on motion data, the quality of which should be sufficient to guide sampling in RL training, as we only use one reference motion. Also, the learned policy has limited generalization in significant changes of setting, \eg strong waves, large fluid densities, strong perturbations, and large changes in body morphology, \etc. In future, we will incorporate lower quality motion data such as those reconstructed from videos to broaden the applicability of SWIM. We will also explore curriculum learning to gradually train the character from simple tasks, a standard body, and quiet water, to progressively harder settings to improve the generalizability.

\section*{Acknowledgements}
This work was supported in part by the UK Research and Innovation AIRR Innovator Award (\mbox{0261-5654-9320-1}), facilities provided by UCL Cosmoparticle Initiative, and the Porto Biomechanics Laboratory (LABIOMEP), University of Porto, and the Qualisys Library for providing the swimming motion dataset.

\newlength{\qualw}
\newcommand{\qfont}{\fontencoding{T1}\fontfamily{phv}\selectfont}  %
\newcommand{\qhead}[3]{\makebox[\linewidth]{\makebox[0.333\linewidth]{\qfont\scriptsize #1}\makebox[0.333\linewidth]{\qfont\scriptsize #2}\makebox[0.333\linewidth]{\qfont\scriptsize #3}}\par\vspace{1pt}}
\newcommand{\qrow}[2]{{\noindent\raggedright\qfont\scriptsize #1\par}\vspace{1pt}\noindent\includegraphics[width=\linewidth]{Figures/qual/#2}\par\vspace{1pt}}
\newcommand{\qcmp}[2]{\rotatebox{90}{\parbox{1.25cm}{\centering\qfont\scriptsize #1}} & \parbox[c]{\qualw}{\includegraphics[width=\qualw]{Figures/qual/#2}}\\[2pt]}
\clearpage
\begin{figure}[p]
\centering
\begin{minipage}{0.91\linewidth}
\includegraphics[width=\linewidth]{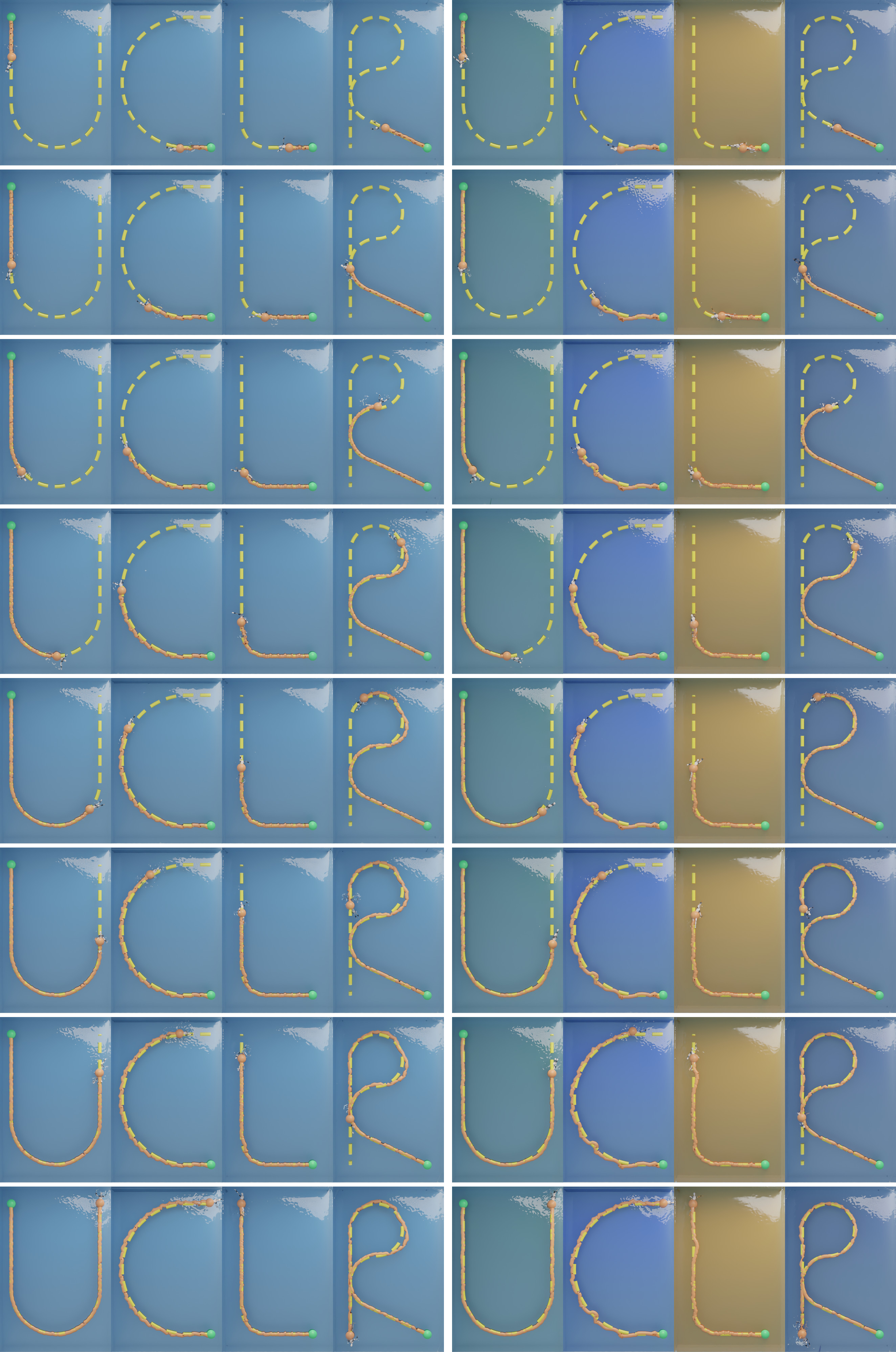}
\end{minipage}
\caption{\textbf{Writing UCLR.} Writing UCL Research ('UCLR'). Four trajectory-following episodes, one per letter: U (butterfly), C (front crawl), L (backstroke), and R (breaststroke). \textbf{Left:} all four letters in still water. \textbf{Right:} U with both lower leg removed, C in a $0.51$\,m/s wave along the pool, L in honey, and R in $0.5\,mg$ perturbation every $6$\,s.}
\label{fig:supp_uclr_word}
\end{figure}

\begin{figure}[p]
\centering
\qrow{(a) Front crawl}{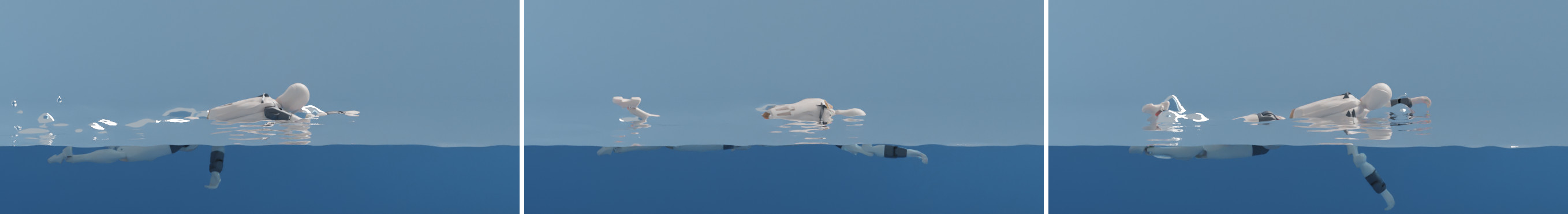}
\qrow{}{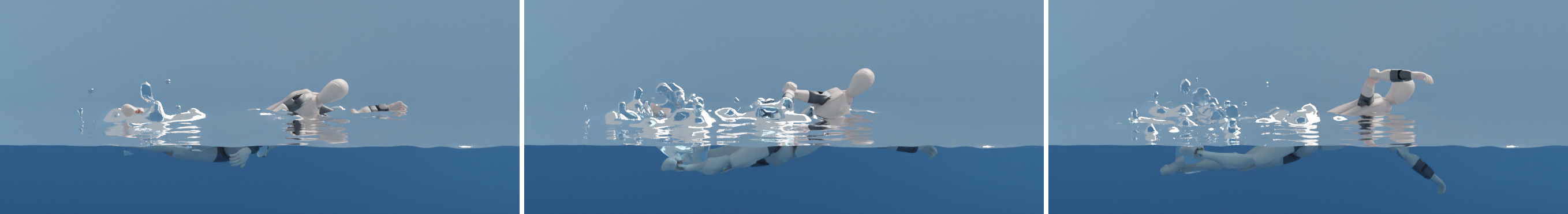}
\qrow{(b) Backstroke}{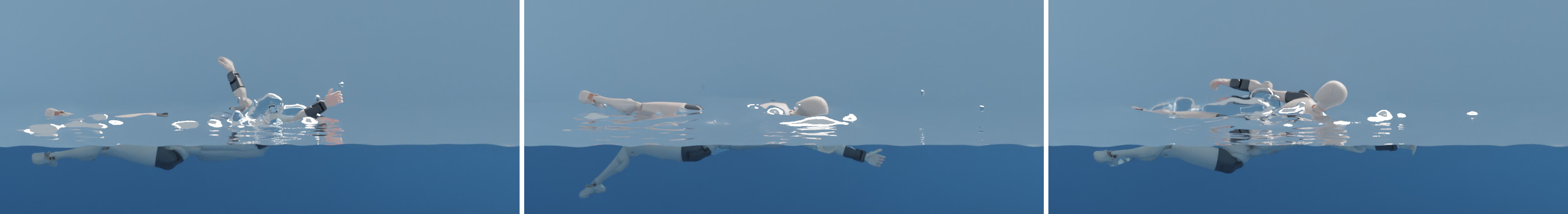}
\qrow{}{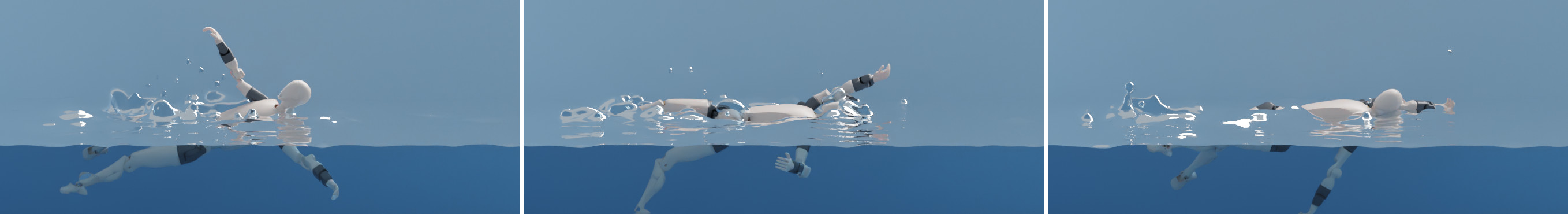}
\qrow{(c) Breaststroke}{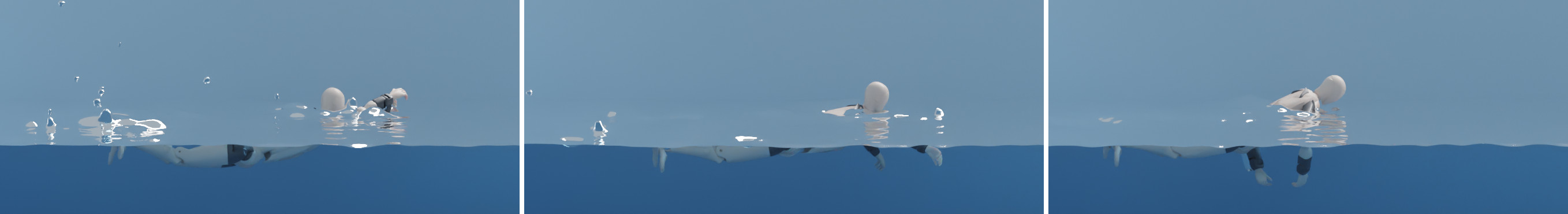}
\qrow{}{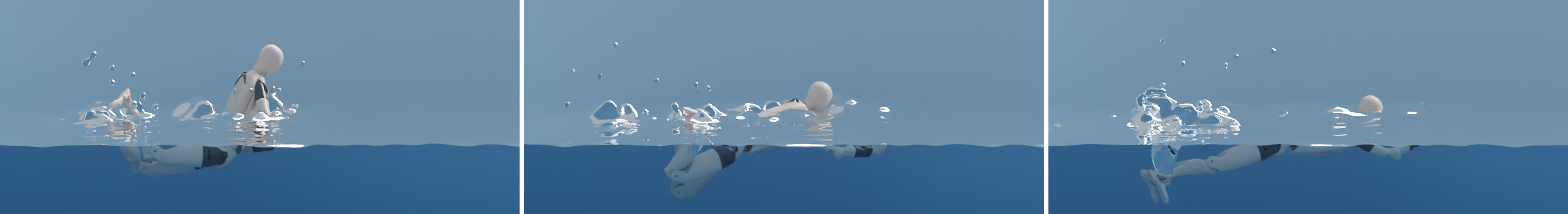}
\qrow{(d) Butterfly}{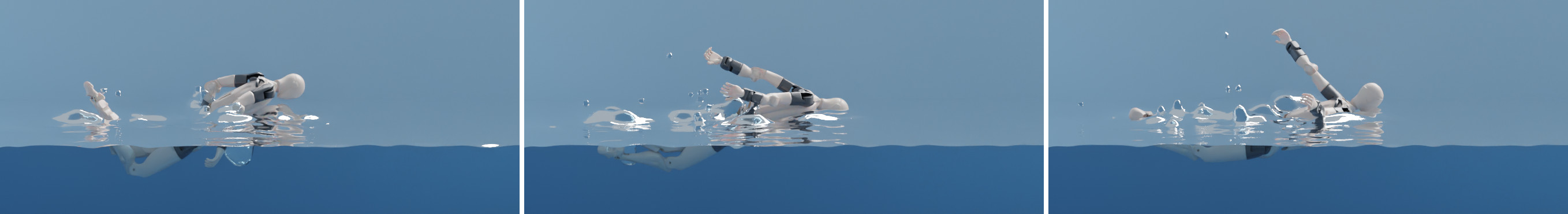}
\qrow{}{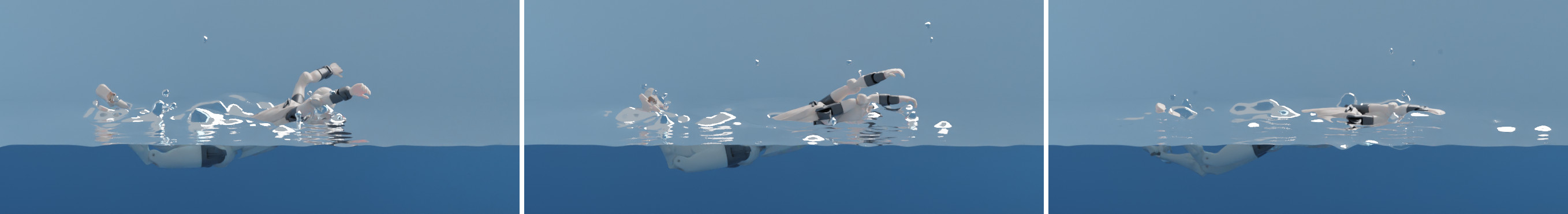}
\vspace{-4pt}
\caption{\textbf{One stroke cycle, four strokes.} Side view on the goal-reaching task in still water. Each stroke takes two rows with six key poses from one cycle, read left to right and top to bottom. The cycle length differs by stroke, from $1.06$\,s for butterfly to $1.58$\,s for front crawl.}
\label{fig:supp_phase_base}
\end{figure}

\begin{figure}[p]
\centering
\qrow{(a) Water flow speed 0.51\,m/s, front crawl}{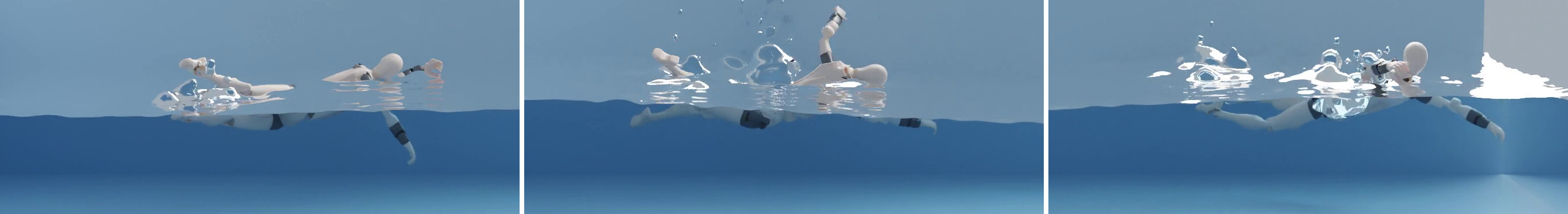}
\qrow{(b) Water flow speed 0.51\,m/s, butterfly}{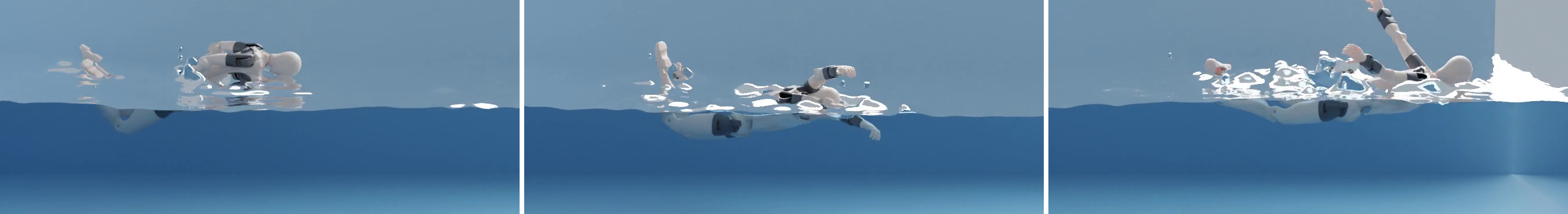}
\qrow{(c) Different fluid: glycerol, breaststroke}{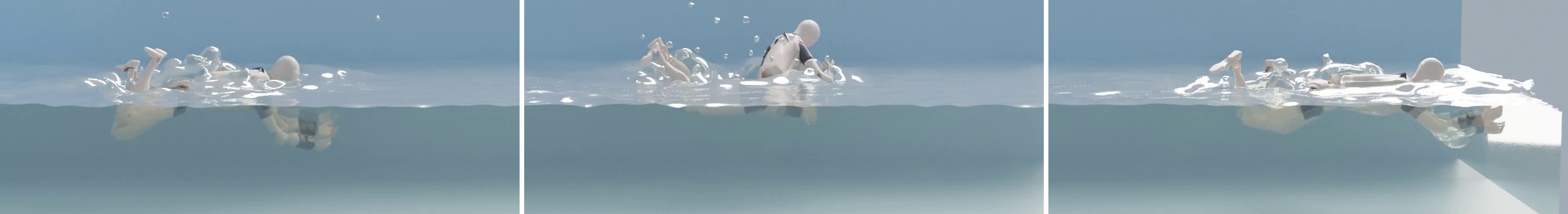}
\qrow{(d) Different fluid: glycerol, backstroke}{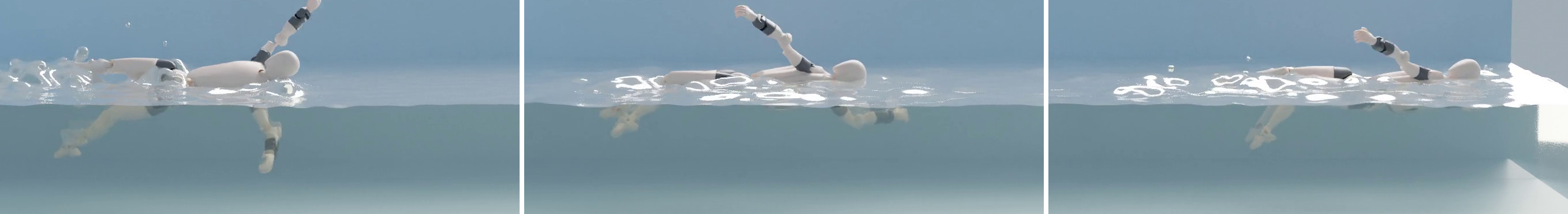}
\qrow{(e) Downward push of 1\,\textit{mg}, butterfly}{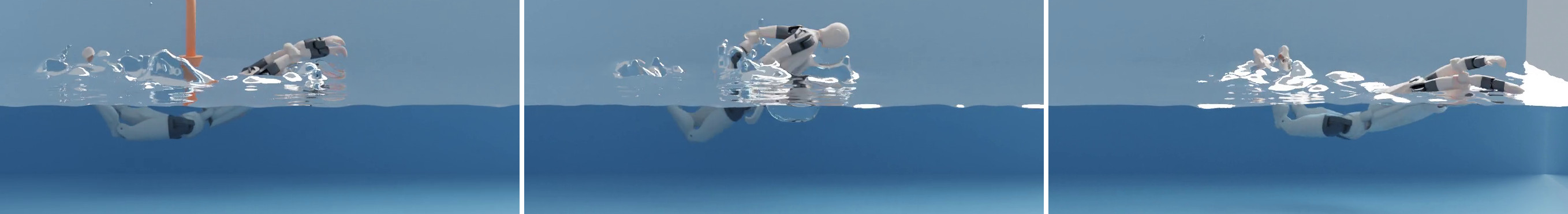}
\qrow{(f) Backward push of 1\,\textit{mg}, breaststroke}{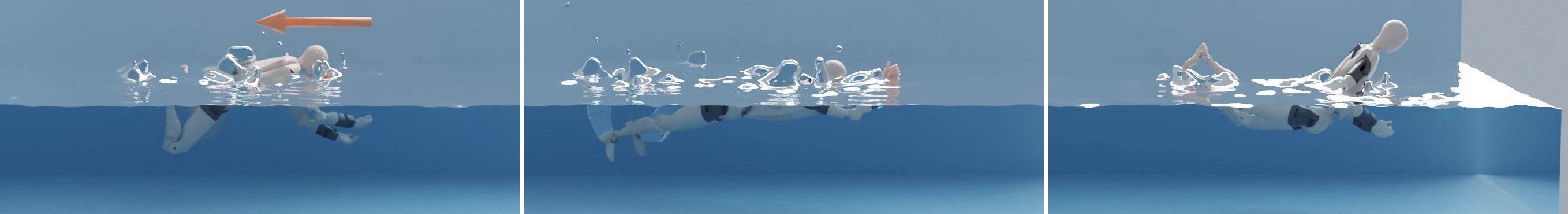}
\qrow{(g) 10\,cm fins, front crawl}{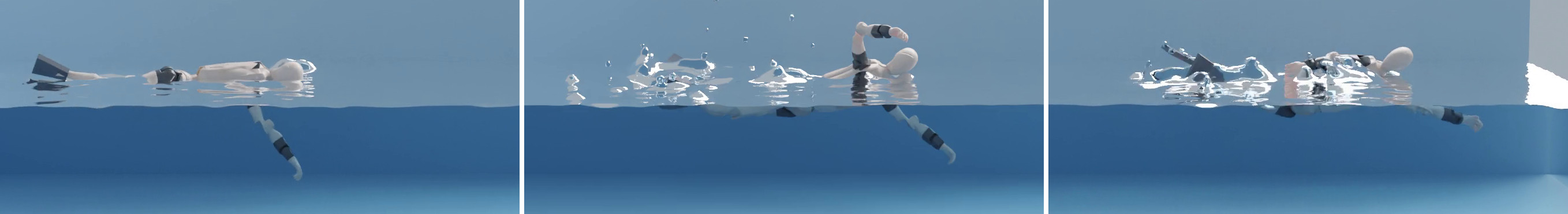}
\qrow{(h) Right leg removed, butterfly}{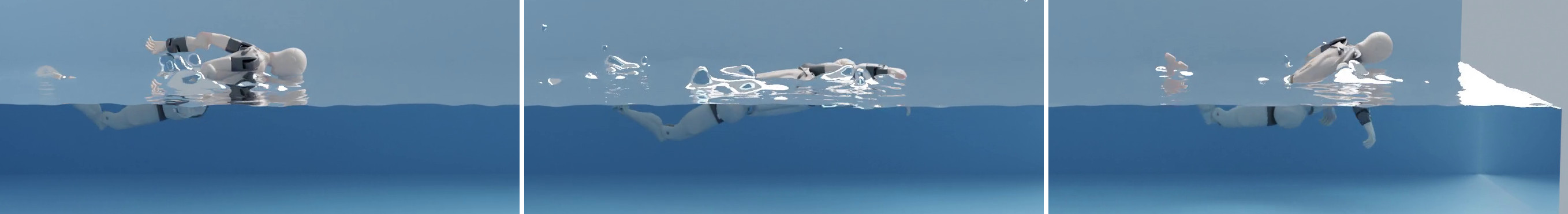}
\vspace{-4pt}
\caption{\textbf{Unseen conditions, side view.} Each row shows one SWIM experiment, with the view following the character. Frames are shown at approximately $2$, $5$, and $8$\,s, except in (e) and (f), where they are captured during the push at approximately $4.5$\,s and then $1$ and $3$\,s later. The orange arrow marks the push, with a force of one body weight ($1\,mg$) applied for $0.1$\,s. The policy maintains control during the push and recovers afterward. Every episode shown reaches its goal.}
\label{fig:supp_gen_side}
\end{figure}

\begin{figure}[p]
\centering
\begin{minipage}{0.80\linewidth}
\qrow{(a) Training: T1 goal-reaching in the 3.5\,m training pool at the three training goals, front crawl}{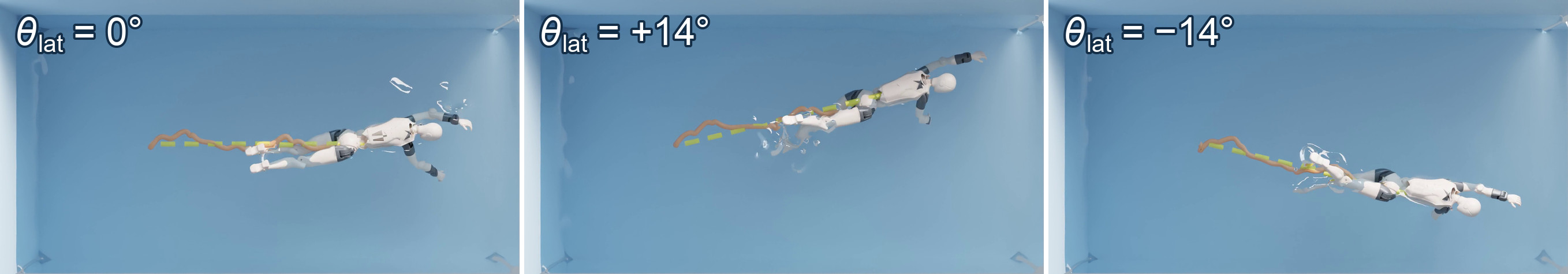}
\qrow{(b) T1 goal-reaching at \textendash{}30\textdegree{} in the 8\,m pool, front crawl}{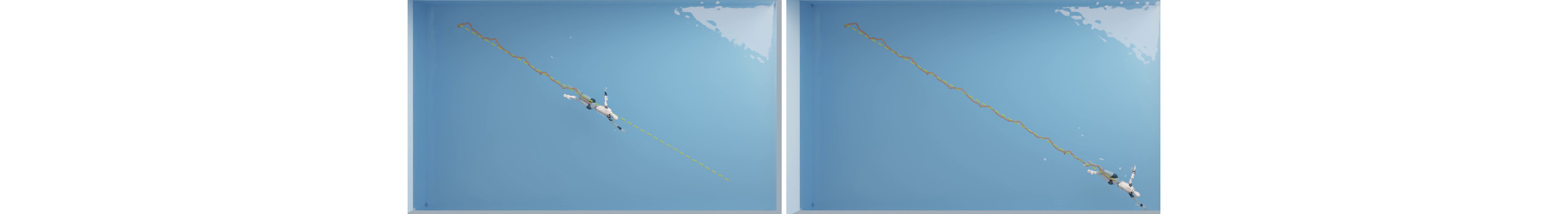}
\qrow{(c) T2 symmetric V at \textendash{}10\textdegree{} in the 8\,m pool, backstroke}{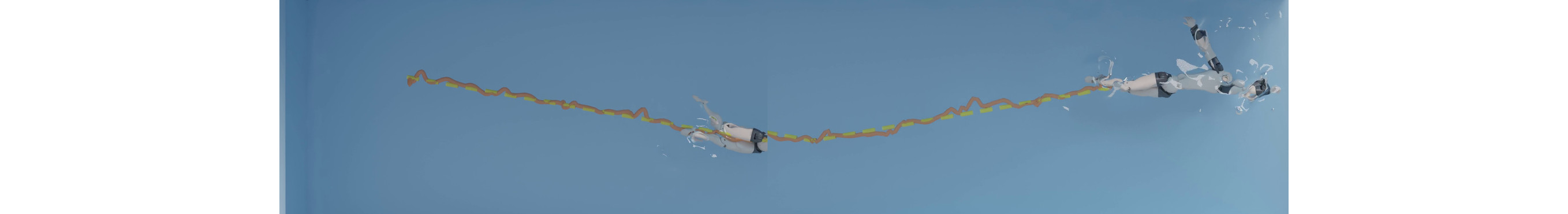}
\qrow{(d) T3 asymmetric V at +15\textdegree{} in the 8\,m pool, breaststroke}{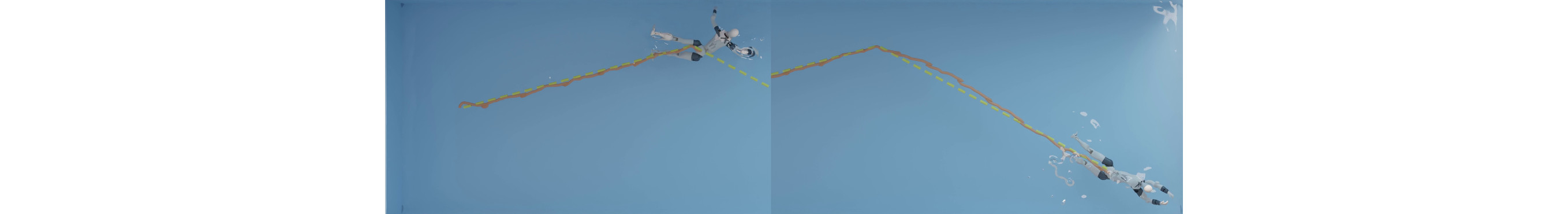}
\qrow{(e) T4 half-sine at +30\textdegree{} in the 8\,m pool, butterfly}{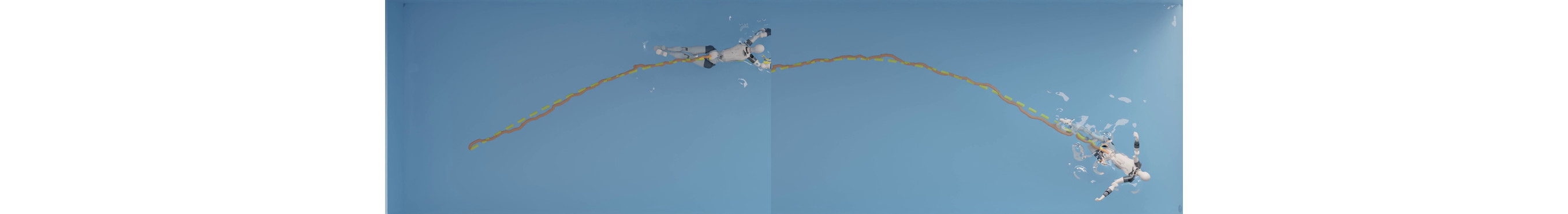}
\qrow{(f) T4 at +30\textdegree{} with upstream flow 0.51\,m/s, front crawl}{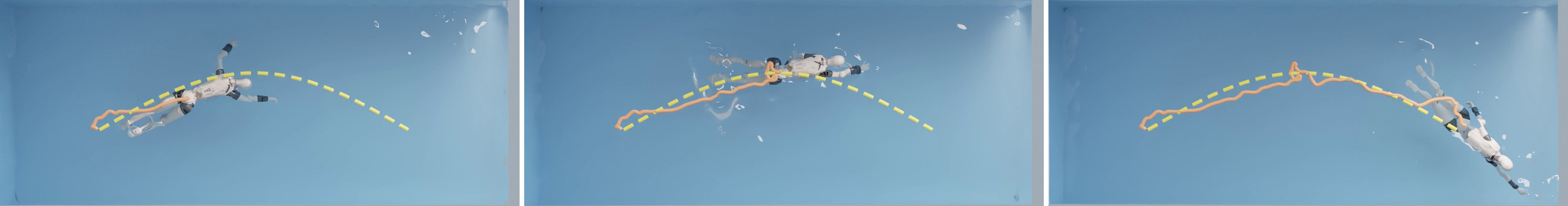}
\qrow{(g) T2 at +15\textdegree{} in glycerol, breaststroke}{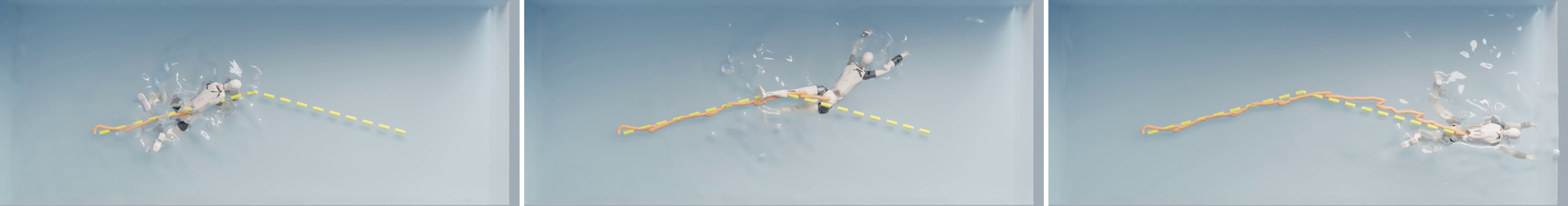}
\qrow{(h) T2 at +15\textdegree{} with a sideways push, butterfly}{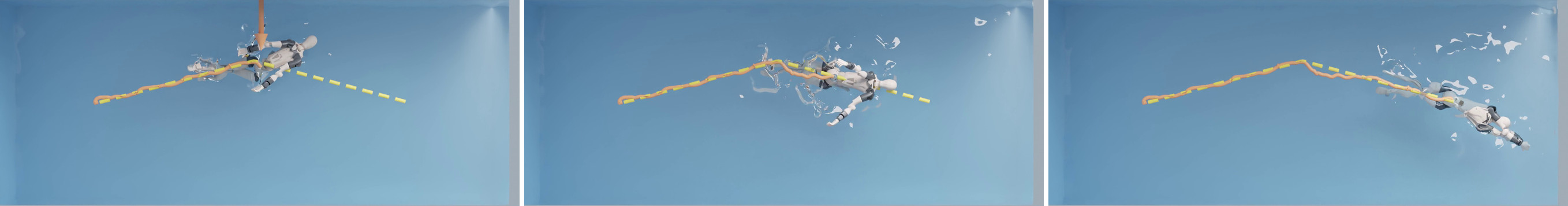}
\qrow{(i) T1 at +4.7\textdegree{} with 10\,cm fins, front crawl}{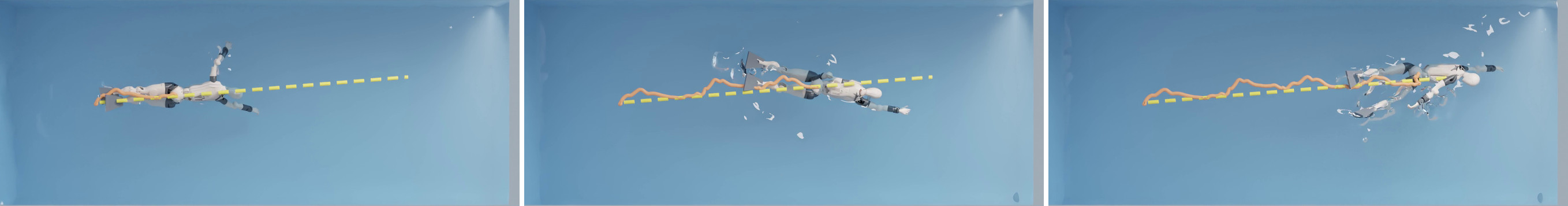}
\qrow{(j) T1 at +4.7\textdegree{} with the right leg removed, butterfly}{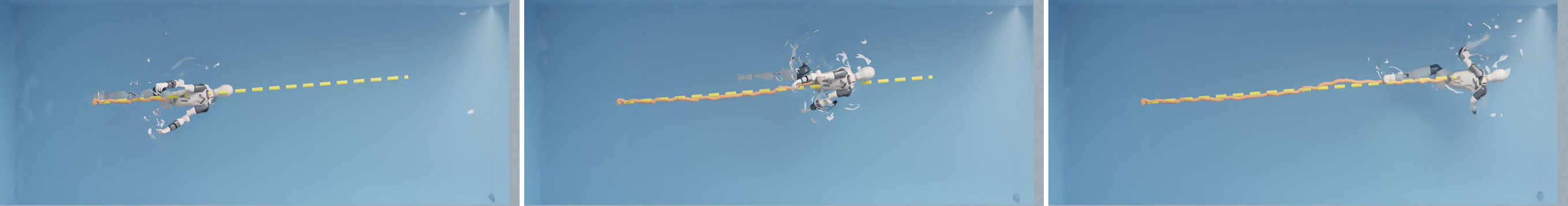}
\end{minipage}
\vspace{-7pt}
\caption{\textbf{Tasks and conditions, top view.} (a): the training task, goal-reaching in still water in the small training pool at the three training goals, one episode each at its end ($4$\,s). The other rows are one SWIM episode each. (b) to (e): the four tasks in still water, one per stroke, in the unseen $8$\,m pool; each shows the whole pool with the swimmer at $8$ and $16$\,s. (f) to (j): one unseen condition each, at about $2$, $5$ and $8$\,s, except (h), which is during the push at about $4.2$\,s, about $1$\,s later and at $8$\,s. Dashed yellow: target path. Orange: swum path. Every episode shown reaches its goal.}
\label{fig:supp_gen_top}
\end{figure}

\begin{figure}[p]
\centering
\setlength{\qualw}{\dimexpr\linewidth-0.55cm\relax}
\setlength{\tabcolsep}{1pt}
\begin{tabular}{@{}cc@{}}
& \makebox[\qualw]{\makebox[0.25\qualw]{\qfont\scriptsize Front crawl}\makebox[0.25\qualw]{\qfont\scriptsize Backstroke}\makebox[0.25\qualw]{\qfont\scriptsize Breaststroke}\makebox[0.25\qualw]{\qfont\scriptsize Butterfly}}\\[1pt]
\qcmp{\textbf{SWIM (ours)}}{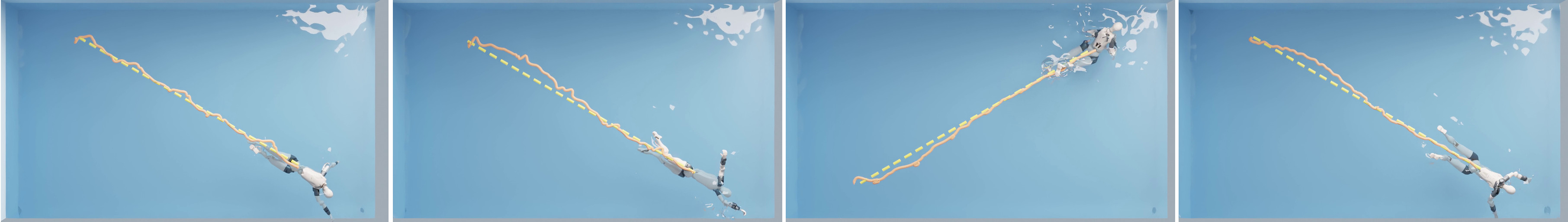}
\qcmp{MimicKit-DeepMimic}{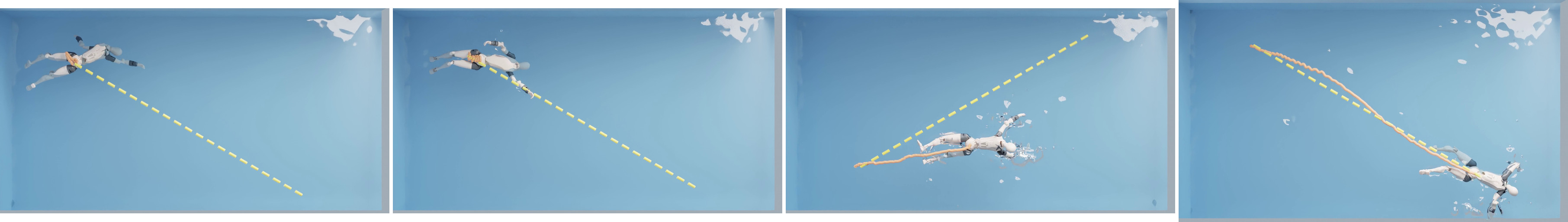}
\qcmp{MimicKit-AMP}{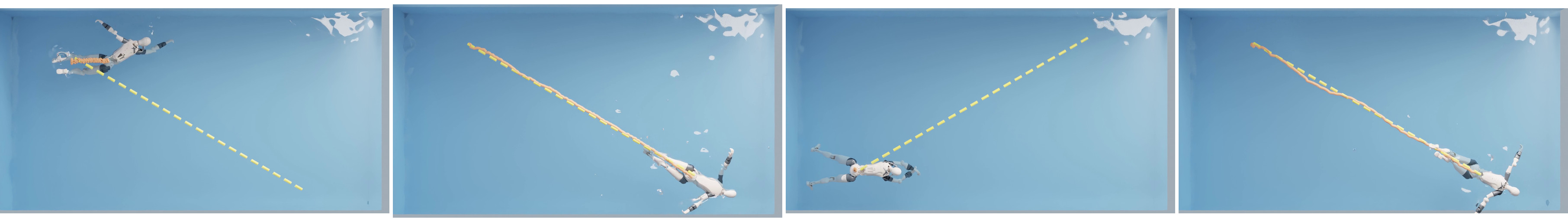}
\qcmp{ADD}{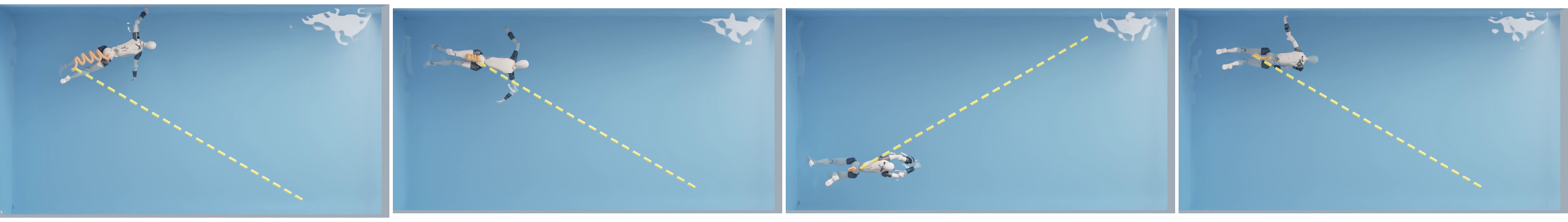}
\qcmp{MuJoCo}{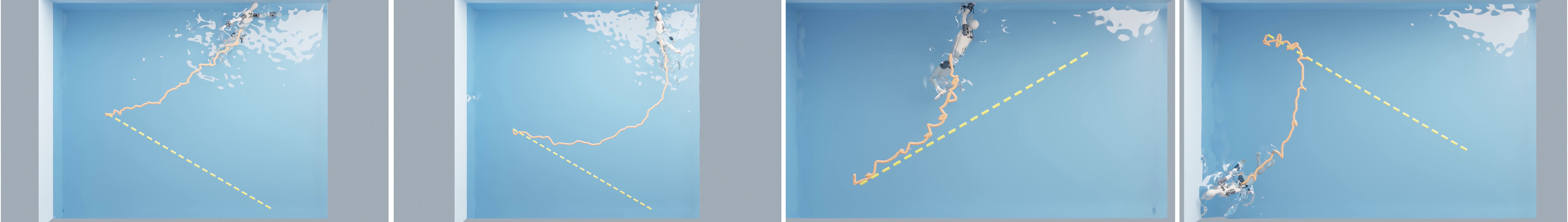}
\qcmp{CFC}{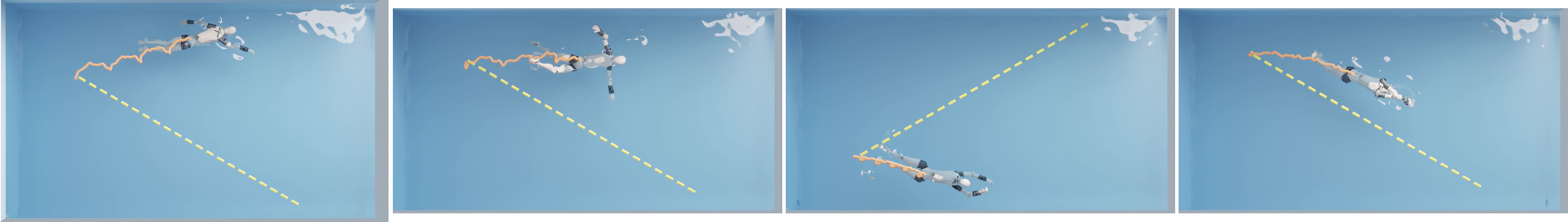}
\qcmp{OceanSim}{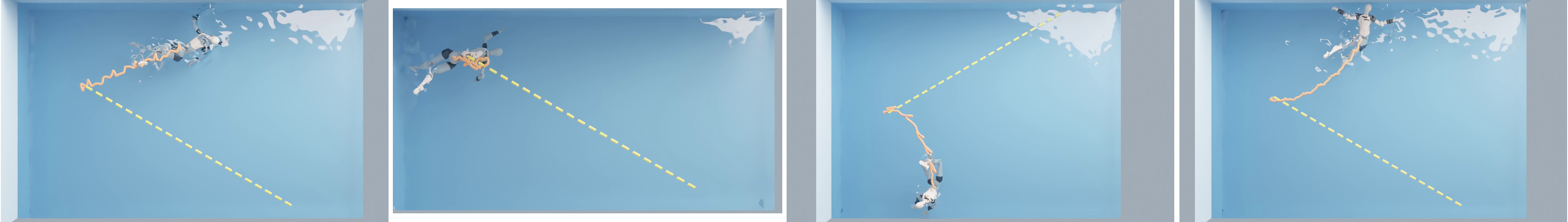}
\qcmp{High-res SPH}{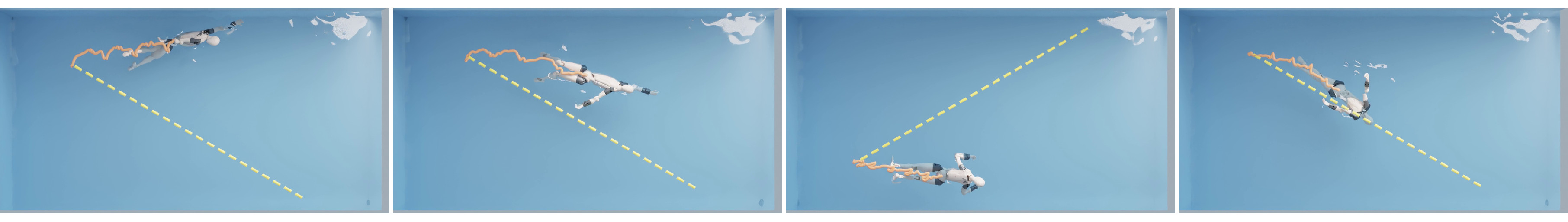}
\end{tabular}
\caption{\textbf{SWIM and the baselines on the same task.} T1 goal-reaching at $30^\circ$ in still water ($-30^\circ$, and $+30^\circ$ for breaststroke), about twice the largest training angle, shown at the end of the episode ($8$\,s). All policies are run in our SPH fluid. Dashed yellow: the straight path to the goal. Orange: the path swum. SWIM reaches the goal in all four strokes, MimicKit-AMP in backstroke and butterfly, and MimicKit-DeepMimic in butterfly; no other method reaches it. Some MuJoCo and OceanSim swimmers leave the lane, so their pool is drawn wider.}
\label{fig:supp_compare}
\end{figure}
\clearpage   %

\begin{table}[tb]
\centering
\small
\setlength{\tabcolsep}{4pt}
\caption{\textbf{Number of experiments.} One experiment is one policy swimming one task under one condition. Rows naming several methods or arms multiply by that count.}
\label{tab:supp_experiment_scale}
\vspace{0.1cm}
\resizebox{\textwidth}{!}{%
\begin{tabular}{l|rrrr|r|r}
\toprule
 & Conditions & Tasks & Strokes & Seeds & Experiments & Subtotal \\
\midrule
\multicolumn{7}{l}{\emph{Strong generalization} (\Cref{sec:generalisation})} \\
\midrule
$\mathbf{T}\times\mathbf{E1}$ (flow) & 8 & 30 & 4 & 1 & 960 & \multirow{8}{*}{16{,}632} \\
$\mathbf{T}\times\mathbf{E2}$ (fluid) & 8 & 30 & 4 & 1 & 960 & \\
$\mathbf{T}\times\mathbf{P}$ (perturbation) & 91 & 30 & 4 & 1 & 10{,}920 & \\
$\mathbf{T}\times\mathbf{B}$ (body) & 11 & 30 & 4 & 1 & 1{,}320 & \\
$\mathbf{T}$ in the $4\times$ pool & 1 & 30 & 4 & 1 & 120 & \\
Goal-reaching angle sweep, $\pm5^\circ$ to $\pm45^\circ$ & 1 & 18 & 4 & 1 & 72 & \\
$\mathbf{T}\times$ additional flow speeds & 16 & 30 & 4 & 1 & 1{,}920 & \\
$\mathbf{T}\times$ additional densities & 3 & 30 & 4 & 1 & 360 & \\
\midrule
\multicolumn{7}{l}{\emph{Baselines} (\Cref{tab:baselines}), on $\mathbf{T}\times\mathbf{E1}$} \\
\midrule
Imitation baselines (MimicKit-AMP, MimicKit-DeepMimic, ADD) & 8 & 30 & 4 & 3 & 8{,}640 & \multirow{2}{*}{20{,}160} \\
Transition-function baselines (MuJoCo, CFC, OceanSim, high-res SPH) & 8 & 30 & 4 & 3 & 11{,}520 & \\
\midrule
\multicolumn{7}{l}{\emph{Ablation study} (\Cref{tab:ablation_overall} and the supplementary material), on $\mathbf{T}\times\mathbf{E1}$} \\
\midrule
Environment state (8 arms) & 8 & 30 & 4 & 3 & 23{,}040 & \multirow{2}{*}{46{,}080} \\
Reward, goal state and action space (8 arms) & 8 & 30 & 4 & 3 & 23{,}040 & \\
\midrule
\textbf{Total} & & & & & & \textbf{82{,}872} \\
\bottomrule
\end{tabular}}%
\end{table}

\begin{table}[tb]
\centering
\small
\setlength{\tabcolsep}{4pt}
\caption{\textbf{Number of trained policies.} Each method is trained from scratch with three seeds for each of the four strokes. The generalization experiments in \Cref{tab:supp_experiment_scale} reuse the best seed for each stroke without any additional training.}
\label{tab:supp_policy_count}
\vspace{0.1cm}
\resizebox{\textwidth}{!}{%
\begin{tabular}{l|rrr|r}
\toprule
 & Methods  & Strokes & Seeds & Policies \\
\midrule
Environment state & 8 & 4 & 3 & 96 \\
Reward, goal state and action space & 8 & 4 & 3 & 96 \\
Imitation baselines (MimicKit-AMP, MimicKit-DeepMimic, ADD) & 3 & 4 & 3 & 36 \\
Transition-function baselines (MuJoCo, CFC, OceanSim, high-res SPH) & 4 & 4 & 3 & 48 \\
\midrule
\textbf{Total} & & & & \textbf{276} \\
\bottomrule
\end{tabular}}%
\end{table}

\clearpage

\bibliography{references}
\bibliographystyle{swim_preprint}

\end{document}